\documentclass[final,3p,times]{elsarticle} 

\usepackage{hyperref}
\usepackage{xcolor}
\usepackage{algorithm}
\usepackage{algpseudocode}
\usepackage{siunitx}
\usepackage{mathtools}
\usepackage{cases}
\ifdefined\qtyproduct
\else
  \ifdefined\NewCommandCopy
    \NewCommandCopy\qtyproduct\SI
  \else
    \NewDocumentCommand\qtyproduct{O{}mm}{\SI[#1]{#2}{#3}}
  \fi
\fi

\usepackage{lipsum}
\newcommand\blfootnote[1]{%
  \begingroup
  \renewcommand\thefootnote{}\footnote{#1}%
  \addtocounter{footnote}{-1}%
  \endgroup
}

\journal{Computers and Structures}
\everymath{\displaystyle}

\begin{document}

\blfootnote{This manuscript version is made available under the CC-BY-NC-ND 4.0 license.}
\blfootnote{DOI:~\url{https://doi.org/10.1016/j.compstruc.2022.106862}}

\begin{frontmatter}


\title{A parallel algorithm for unilateral contact problems}


\author[bsc]{G. Guillamet\corref{cor1}}\ead{gerard.guillamet@bsc.es}
\author[bsc]{M. Rivero}
\author[bsc]{M. Zavala-Aké}
\author[bsc]{M. Vázquez}
\author[bsc]{G. Houzeaux}
\author[upc,cim]{S. Oller}
\cortext[cor1]{Corresponding author}
\address[bsc]{Barcelona Supercomputing Center (BSC), Plaça Eusebi Güell, 1-3, 08034 Barcelona, Spain}
\address[upc]{Universitat Politècnica de Catalunya (UPC), Campus Nord UPC, 08034 Barcelona, Spain}
\address[cim]{Centre Internacional de Mètodes Numèrics en Enginyeria (CIMNE), Campus Nord UPC, 08034 Barcelona, Spain}

\begin{abstract}
In this paper, we introduce a novel parallel contact algorithm designed to run efficiently in High Performance Computing based supercomputers. Particular emphasis is put on its computational implementation in a multiphysics finite element code. The algorithm is based on the method of partial Dirichlet-Neumann boundary conditions and is capable to solve numerically a nonlinear contact problem between rigid and deformable bodies in a whole parallel framework. Its distinctive characteristic is that the contact is tackled as a coupled problem, in which the contacting bodies are treated separately, in a staggered way. Then, the coupling is performed through the exchange of boundary conditions at the contact interface following a Gauss-Seidel strategy.
To validate this algorithm we conducted several benchmark tests by comparing the proposed solution against theoretical and other numerical solutions. Finally, we evaluated the parallel performance of the proposed algorithm in a real impact test to show its capabilities for large-scale applications.
\end{abstract}


\end{frontmatter}

\section{Introduction}\label{sec:intro}

Since decades ago, contact problems have taken an important place in computational solid mechanics. Because of their relevance and complexity, many numerical procedures have been proposed in the engineering literature. The pioneering works to treat contact mechanics within the Finite Element Method were done by Francavilla and
Zienkiewicz~\citep{Francavilla1975} and Hughes~\cite{Hughes1976}. In these works, they proposed a simple node-based approach that requires node-matching meshes between contacting bodies and are restricted to small deformations. Since then, the field of computational contact mechanics has experienced a continuous evolution. An example of this are the methods proposed by Laursen~\cite{Laursen2003}, Wriggers and Zavarise~\cite{Wriggers2006} or more recently by Yastrebov~\cite{Yastrebov2013} for the numerical treatment of contact problems.

The most used methods within finite element framework are the ones with constraint enforcement such as the penalty method~\cite{Hallquist1985,Simo1985,Wriggers1990} and the classical Lagrange multiplier method~\cite{Bathe1985,Gallego1989a}. Due to known drawbacks of the penalty method (see~\cite{Kikuchi1988}), Lagrange multiplier techniques have become relevant in the domain of constrained problems. Especially one of its extensions has gained importance in the field of computational contact mechanics: the Augmented Lagrangian method, which combines advantages of both penalty and Lagrange multipliers and has been applied successfully to frictionless and frictional contact (see~\cite{Glowinski1989,Pietrzak1999,Simo1992}). 

With respect to the discretization of the continuum setting, these approaches result in a monolithic system of linear equations which includes all the unknowns for all the geometries of the mechanical system. Furthermore, the discretization of contact problems using implicit solvers requires the construction of contact elements, contact tangent matrices, and contact residual vectors which incorporate the contact constraints in the global weak form. In almost all contact problems where the contact zone is a priori unknown, the use of contact elements requires the update of the mesh graph in a fixed number of time steps. On the other hand, some of the optimization techniques used in contact problems increase the number of degrees of freedom in the problem, by introducing Lagrange multipliers as unknowns. Since the number of Lagrange multipliers depends on the active contact surface at each time step, the total unknowns of the system can vary as the problem evolves. 

As an alternative to the previous methods, there are formulations that allow solving problems for each body separately, with certain boundary conditions at the natural interface. These kinds of methods are named as Dirichlet-Neumann type contact algorithms, and for a complete overview, the reader is referred to~\cite{Krause2002,Bayada2002,Haslinger2009,YastrebovPhd2011,Eck2003,Haslinger2014}. One of the pioneering works in this area was done by Krause~\cite{Krause2002}, who solved a frictional Signorini contact problem between two elastic bodies based on mortar elements for discretization, and a monotone multigrid method as a subdomain solver. The main idea of this algorithm is to solve in each iteration a linear Neumann problem for one body and a unilateral contact problem for the other body, by using essentially the contact interface as the boundary data transfer.  Convergence of this algorithm in the continuous setting in its frictionless form has been proved in~\cite{Bayada2002,Eck2003} and considering friction in~\cite{Bayada2008}. More recently, Yastrebov~\cite{YastrebovPhd2011,Yastrebov2011b} proposed a method for unilateral frictional contact in which the contact constraints are replaced by partial Dirichlet-Neumann boundary conditions: normal contact is introduced as a Dirichlet condition and frictional contact is introduced as a Neumann condition. One of the most attractive features of this algorithm is its simplicity and stability in the fulfillment of the geometrical constraints. 

As parallel computing platforms are nowadays widely available, solving large engineering problems on high-performance computers in supercomputing facilities is a possibility for any engineer or researcher. Due to the memory and computing power that contact problems require and consume, they are good candidates for parallel computation. Industrial and scientific realistic contact problems involve different physical domains and a large number of degrees of freedom, so algorithms designed to run efficiently in high-performance computers are needed. Nevertheless, the parallelization of the solution methods which arises from the classical optimization techniques and discretization approaches for the solution of contact problems present some drawbacks and difficulties. Few works in the computational contact mechanics literature have been devoted to parallel methods, also reporting their difficulty and effort to parallelize this kind of problems~\cite{Malone1994a,Attaway1998,Har2003,Yastrebov2011b}. In the finite element formulation, the contact element matrices are assembled in the global structural matrix of the system. Since traditional contact formulations are based on contact elements that create additional node connectivities, from a parallelization point of view this presents some drawbacks when the contact area changes during execution time. Sliding is a very frequent phenomenon in contact problems that requires the modification of node connectivities at the boundary contact zone due to the creation and destruction of contact elements. Since the contact area changes during an incremental solution procedure, the mesh graph must be updated in a fixed number of time steps. From the point of view of the parallel resolution, this is a significant drawback due to its computational expensiveness. A dynamic repartitioning must be done to redistribute the updated mesh graph to the different processors. On the other hand, some contact methods use Lagrange multipliers as unknowns, which also affects the global structural matrix by dynamically modifying the number of degrees of freedom as the contact area changes. 

Motivated by the fact that standard contact algorithms are not a suitable alternative for efficient parallelization, we introduce a novel contact algorithm that can run efficiently in High Performance Computing (HPC) based supercomputers. In this paper, we put special emphasis on its computational implementation. This algorithm, which is based on the method of partial Dirichlet-Neumann boundary conditions (PDN) proposed by Yastrebov~\cite{YastrebovPhd2011}, is designed to solve nonlinear contact problems between rigid-to-deformable bodies in a whole parallel framework. One of the main distinctive characteristics of this algorithm is that the contact is tackled as a coupled problem. As a consequence, the bodies in contact are treated separately, in a staggered way. Then, the coupling is performed through the exchange of boundary conditions at the contact interface, following a Gauss-Seidel strategy. As this approach solves each body separately, there is no need to increase the degrees of freedom of the problem or to redefine the mesh graph at different time steps, since no contact elements and no Lagrange multipliers are used. The mesh partitioning is only done at the beginning of the execution, as a preprocess task, without restricting the partitioning algorithm. The contact detection is based on an Eulerian-Lagrangian system analogy, in which the geometry of the rigid body is used as a reference for the detection of contact. Due to the particular characteristics that this algorithm presents, it can be interpreted as a black-box scheme that can be used with any parallel finite element code. All the reasons enumerated previously make this algorithm suitable for solving large-scale contact problems on parallel computers. 

The rest of this paper is outlined as follows: Section~\ref{sec:formulationunilateral} summarizes the formulation for unilateral contact problems. Section~\ref{sec:pdnmethod} describes the method of partial Dirichlet-Neumann which is the basis of the proposed algorithm. Then, Section~\ref{sec:computimp} describes the computational implementation and the parallel aspects which are required for an HPC finite element code. Section~\ref{sec:benchmark} includes a series of benchmark problems for the verification and validation of the proposed algorithm. Some of these benchmark problems are used to evaluate the computational performance of the proposed algorithm in Section~\ref{sec:parallperformance}. Finally, the conclusions of the present work are summarized at the end of this article.

\section{Formulation of unilateral contact problems}\label{sec:formulationunilateral}

\subsection{Unilateral normal contact}

\begin{figure}[h]
\begin{center}
\includegraphics[width=8.5cm]{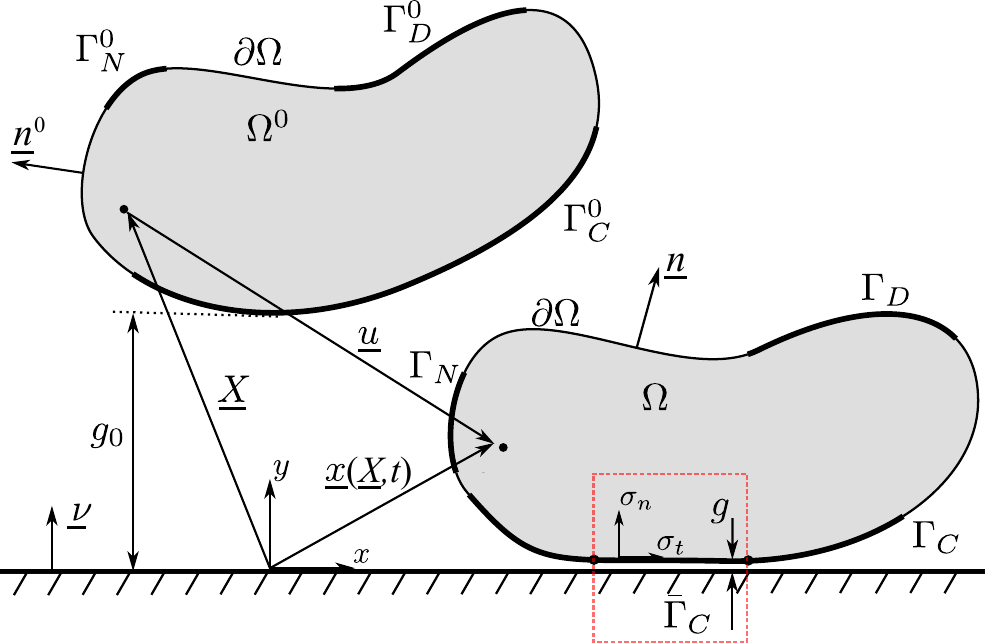} 
\caption{Reference ($\Omega_0$) and current ($\Omega$) configurations of a deformable body in unilateral contact with a rigid plane.}
\label{fig:scheme1}
\end{center}
\end{figure}

Let's assume the motion of a deformable body approaching to a rigid plane, see Fig.~\ref{fig:scheme1}. The motion is described with respect to a fixed orthonormal coordinate system with basis vectors \{$\underline{e_{x}},\underline{e_{y}}$\}. All the points that belong to the body are described by a vector $\underline{X}$ in the reference configuration $(\Omega^{0})$ and by a vector $\underline{x}(\underline{X},t)$ in the current configuration $(\Omega)$ at time $t$. Being the rigid plane in this position, the motion of any body in space is then restricted to $y \geq 0$ from the rigid plane. Then, we can represent this restriction by the following displacement contact constraint:

\begin{equation}\label{eq:gap1}
g(\underline{\boldsymbol{x}}) = \underline{\boldsymbol{x}} \cdot \underline{\boldsymbol{\nu}} \geq 0,
\end{equation}

where $g(\underline{\boldsymbol{x}})$ is the gap between the current point $\underline{\boldsymbol{x}}$ of the body and the rigid plane and $\underline{\boldsymbol{\nu}} = \underline{\boldsymbol{e}}_n$ is a unit vector normal to the rigid plane pointing to the area of motion. In other words, Eq.~\eqref{eq:gap1} states that any point of the body at any time can not penetrate the rigid plane.

Defining the displacement of any point of the body as $\underline{\boldsymbol{u}} = \underline{\boldsymbol{x}} - \underline{\boldsymbol{X}}$, we can express Eq.~\eqref{eq:gap1} as:

\begin{equation}\label{eq:gap2}
g(\underline{\boldsymbol{u}}) = \underline{\boldsymbol{u}} \cdot \underline{\boldsymbol{\nu}} + g_0 \geq 0,
\end{equation}

where $g_0 = \underline{\boldsymbol{X}} \cdot \underline{\boldsymbol{\nu}}$ is the initial gap. If the body retains its integrity, the non-penetration condition given by Eq.~\eqref{eq:gap2} is applied only to the surface points $\partial \Omega$, precisely to the {\it potential contact zone} ($\Gamma_c$) in the current configuration. $\Gamma_c$ can be splitted into two nonintersecting sets: the active contact zone $\bar{\Gamma}_c$ (points which are in contact) and the inactive contact zone $\Gamma_c \backslash \bar{\Gamma}_c$ (points which are not in contact). The active contact zone in the actual configuration is defined by:

\begin{equation}\label{eq:activecontactzone1}
\underline{\boldsymbol{x}} \in \bar{\Gamma}_c \quad \textrm{if and only if} \quad g(\underline{\boldsymbol{x}}) = \underline{\boldsymbol{x}} \cdot \underline{\boldsymbol{\nu}} = 0,
\end{equation}

while for the reference configuration, the active contact zone can be defined by:

\begin{equation}\label{eq:activecontactzone2}
\underline{\boldsymbol{X}} \in \bar{\Gamma}^0_c \quad \textrm{if and only if} \quad  \underline{\boldsymbol{X}} \cdot \underline{\boldsymbol{\nu}} = - \underline{\boldsymbol{u}} \cdot \underline{\boldsymbol{\nu}}.
\end{equation}

From the definitions in Eq.~\eqref{eq:activecontactzone1} and Eq.~\eqref{eq:activecontactzone2} we observe that the active and inactive contact zones are a priori unknowns of the problem. Only in some specific problems, given the potential zone we can predict a priori the active contact zone at each instant of time.

When the contact between the body and the rigid plane occurs, a contact pressure appears in the active contact zone in order to prevent the penetration. This pressure should be zero in the inactive contact zone and negative in the active contact zone:

\begin{equation}\label{eq:contactpressure}
\sigma_n \leq 0 \quad \textrm{at} \quad \Gamma_c.
\end{equation}

Combining the non-penetration condition in the active zone given by Eq.~\eqref{eq:activecontactzone1} and the definition of the contact pressure by Eq. \eqref{eq:contactpressure} we get the non-penetration-non-adhesion condition:

\begin{equation}\label{eq:nonpenenonade}
\sigma_n \, g(\underline{\boldsymbol{x}}) = 0 \quad \textrm{at} \quad \Gamma_c.
\end{equation}

All together, the set of conditions expressed in Eq.~\eqref{eq:gap1}, Eq.~\eqref{eq:contactpressure} and Eq.~\eqref{eq:nonpenenonade} form the Hertz-Signorini-Moreau law of unilateral normal contact:

\begin{equation}\label{eq:hertzsignorinimoreau}
g \geq 0, \,\, \sigma_n \leq 0, \,\, \sigma_n \, g = 0.
\end{equation}

\subsection{Balance of momentum including contact}

The nonlinear contact problem can be written as a boundary value problem, given the following equilibrium condition in $\Omega$ and boundary conditions on $\partial\Omega$, which includes the Hertz-Signorini-Moreau law for normal contact:

\begin{equation}\label{eq:bvp}
\begin{split}
& \nabla \cdot \underline{\underline{\boldsymbol{\sigma}}} + \underline{\boldsymbol{f}}_v = 0 \quad \textrm{in} \,\, \Omega\\
& \underline{\underline{\boldsymbol{\sigma}}} \cdot \underline{\boldsymbol{n}} = \underline{\boldsymbol{\sigma}}_0 \quad \textrm{on} \,\, \Gamma_N\\
& \underline{\boldsymbol{u}} = \underline{\boldsymbol{u}}_0 \quad \textrm{on} \,\, \Gamma_D \\
& g \geq 0, \,\, \sigma_n \leq 0, \,\, \sigma_n\,g = 0, \,\, \underline{\boldsymbol{\sigma}}_t = 0  \quad \textrm{on} \,\, \Gamma_C
\end{split}
\end{equation}

being $\underline{\underline{\boldsymbol{\sigma}}}$ the Cauchy stress tensor, $\underline{\boldsymbol{f}}_v$ a vector of volumetric forces, $\underline{\boldsymbol{\sigma}}_0$ a set of prescribed tractions and $\underline{\boldsymbol{u}}_0$ a set of prescribed displacements. Over $\Gamma_C$ we have imposed the contact boundary conditions: $g$ represents the gap between contacting bodies, $\sigma_n$ is the normal contact pressure and $\underline{\boldsymbol{\sigma}}_t$ is the tangential stress. The tangential stress equal to zero ($\underline{\boldsymbol{\sigma}}_t = 0$) in Eq.~\eqref{eq:bvp} characterizes a frictionless contact case.

\subsection{Interpretation of contact Hertz-Signorini-Moreau conditions}

A contact problem can be directly interpreted as finding the active contact zone and the contact pressure which has to be applied in order to fulfill the contact constraints. However, the problem can be also interpreted from another point of view: instead of prescribing the pressure at the active contact zone, we can impose a displacement according to the contact constraints. In what follows, and without loss of generality, we assume a frictionless case. 

The set of normal contact conditions expressed by Eq.~\eqref{eq:hertzsignorinimoreau} can be separated into two parts for active $\bar{\Gamma}_c$ and inactive $\Gamma_c \backslash \bar{\Gamma}_c$ contact zones:

\begin{subnumcases}{}
g = 0,\,\sigma_n < 0,\,\underline{\boldsymbol{\sigma}}_t=0 & at $\bar{\Gamma}_c$ \label{eq:inter_active} \\
g > 0,\,\sigma_n = 0,\,\underline{\boldsymbol{\sigma}}_t=0 & at $\Gamma_c \backslash \bar{\Gamma}_c$ \label{eq:inter_inactive}  
\end{subnumcases}

According to Eq.~\eqref{eq:gap2} and the definition of the active contact zone given by Eq.~\eqref{eq:activecontactzone2}, the first term of Eq.~\eqref{eq:inter_active} can be written as follows:

\begin{equation}\label{eq:gaptodirichlet}
g = 0 \iff \underline{\boldsymbol{\nu}} \cdot \underline{\boldsymbol{u}} = -g^0 \iff u_n = -g^0
\end{equation} 

Eq.~\eqref{eq:gaptodirichlet} shows that the no-penetration condition represented by Eq.~\eqref{eq:gap2} can be interpreted as a Dirichlet boundary condition. In contrast, the condition represented by Eq.~\eqref{eq:inter_inactive} can be interpreted as a singular Neumann boundary condition $\underline{\boldsymbol{\sigma}}_t = 0$ (free boundary) for the inactive zone. Considering the previous interpretations, we can rewrite the Hertz-Signorini-Moreau conditions of Eq.~\eqref{eq:inter_active} and Eq.~\eqref{eq:inter_inactive} as:

\begin{subnumcases}{}
u_n = -g^0,\,\underline{\boldsymbol{\sigma}}_t = 0 & for $\underline{\boldsymbol{x}} \in \bar{\Gamma}_c$ \label{eq:inter_active2} \\
\underline{\boldsymbol{\sigma}} = 0  & for $\underline{\boldsymbol{x}} \in \Gamma_c \backslash \bar{\Gamma}_c$ \label{eq:inter_inactive2}
\end{subnumcases}

Even more, we can rewrite Eq.~\eqref{eq:inter_active2} using the condition given by Eq.~\eqref{eq:contactpressure} in the definition for the active zone:

\begin{equation}\label{eq:inter_active3} 
u_n = -g^0, \,\, \underline{\boldsymbol{\sigma}}_t = 0 \,\,\, \textrm{for} \,\,\, \{ \underline{\boldsymbol{x}} \,\, | \,\, \underline{\boldsymbol{x}} \in {\Gamma}_c  \,\,\, \textrm{and} \,\,\, \sigma_n(\boldsymbol{x}) < 0 \}. 
\end{equation}

In Eq.~\eqref{eq:inter_active2} (and Eq.~\eqref{eq:inter_active3}, which is equivalent) we have replaced the contact conditions on the active contact zone  $\bar{\Gamma}_c$ by a partial Dirichlet boundary condition ($u_n = -g^0$) and a partial Neumann boundary condition ($\underline{\boldsymbol{\sigma}}_t = 0$).

Writing Hertz-Signorini-Moreau law as expressed in Eq.~\eqref{eq:hertzsignorinimoreau} in the form of Eqs.~\eqref{eq:inter_active2} and~\eqref{eq:inter_inactive2} gives a better understanding of the normal contact boundary conditions for unilateral contact problems. From a numerical point of view, it is easier to prescribe a displacement and then check the sign of the contact pressure, rather than to prescribe a contact pressure in an unknown active contact zone (which is determined by a zero value of the gap).

\section{Method of partial Dirichlet-Neumann boundary conditions}\label{sec:pdnmethod}

The main idea of this method is to replace the geometrical constraints due to normal and tangential contact by Partial Dirichlet-Neumann (PDN) boundary conditions in the case of unilateral contact with an arbitrary rigid surface~\cite{YastrebovPhd2011}. In particular, the geometrical constraints due to normal contact are imposed by means of Multi-Point Constraints (MPC). From a geometrical point of view, MPC can be interpreted as Dirichlet boundary conditions that allow sliding of the contacting node only in the tangential plane, assuming a frictionless case. For the enforcement of MPC a chosen degree of freedom of each contacting node --{\it slave} dof $u_s$-- is written as a linear combination of the other dofs of the same node --{\it master} dofs $u_m^i$, $i = 1,\dots,M$--:

\begin{equation}
u_s = \alpha_i u_m^i + \beta,
\end{equation}

where $\alpha_i$ and $\beta$ are scalar coefficients, and $M$ is the total number of master dofs. The slave dof can be chosen arbitrarily for each contacting node but it is required that $\alpha_i < \infty$. 

\begin{figure}[h]
\begin{center}
\includegraphics[width=8.5cm]{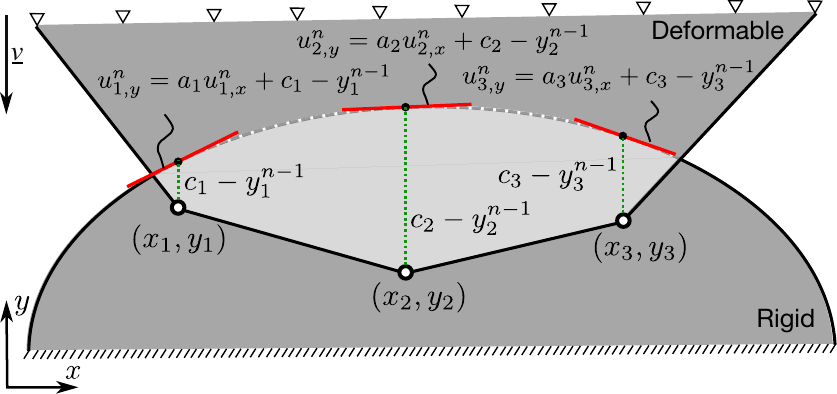} 
\caption{MPC boundary conditions for unilateral frictionless contact.}
\label{fig:MPC1}
\end{center}
\end{figure}

Fig.~\ref{fig:MPC1} shows an example of MPC boundary conditions for a 2D frictionless unilateral contact. The nodes of the deformable body which have penetrated the rigid body are identified and projected following an arbitrary direction to the rigid body surface. Then, a tangent line to the contact surface of the rigid body, which contains the projection point, is computed. This tangent line defines the relation for the MPC boundary condition.

\subsection{Mathematical formulation}\label{sec:pdn_method_math_formul}

Let $\underline{\boldsymbol{x}}^i$ be the coordinates of a contacting node over $\Gamma_c$ in the {\it i}-th iteration. Then, the incremental displacement vector is given by:

\begin{equation}\label{eq:displ}
\underline{\boldsymbol{u}}^i = \underline{\boldsymbol{x}}^i - \underline{\boldsymbol{x}}^{i-1}.
\end{equation} 

The incremental displacement of each degree of freedom is given by splitting the vector $\underline{\boldsymbol{u}}$ into the reference frame basis:

\begin{equation}
u_j^i = \underline{\boldsymbol{u}}^i \cdot \underline{\boldsymbol{e}}_j,
\end{equation}

where $\underline{\boldsymbol{e}}_j$ is a set of basis vectors. 

Otherwise, any rigid surface can be described by the parametric representation:

\begin{equation}
\underline{\mathbf{r}}(s,t) = x(s,t)\,\underline{\boldsymbol{e}}_1 + y(s,t)\,\underline{\boldsymbol{e}}_2 + z(s,t)\,\underline{\boldsymbol{e}}_3.
\end{equation}

Where $s$ and $t$ are the surface parameters. Without any loss of generality, let us suppose that locally exists a function $f$ such that:

\begin{equation}
z = f(x,y),
\end{equation}

where $x$, $y$ and $z$ are coordinates of the surface point in the chosen coordinate system:

\begin{equation}
x = x(s,t)\,\underline{\boldsymbol{e}}_1, \quad y = y(s,t)\,\underline{\boldsymbol{e}}_2, \quad z = z(s,t)\,\underline{\boldsymbol{e}}_3.
\end{equation}

Then, using the definition given by Eq.~\eqref{eq:displ}, the geometrical constraint given by Eq.~\eqref{eq:gap1} can be written in the following way:

\begin{equation}
x_3^i \geq f(x_1^i,x_2^i) \,\,\, \iff \,\,\,  u_3^i \geq f(x_1^{i-1}+u_1^i,x_2^{i-1}+u_2^i) - x_3^{i-1}.
\end{equation}

We can compute the tangential plane at a given point $\{x^*,y^*\}$ on $\Gamma_c$ by the following expression:

\begin{equation}
P : z = \dfrac{\partial f}{\partial x} \Big|_{\{x^*,y^*\}} (x-x^*) + \dfrac{\partial f}{\partial y} \Big|_{\{x^*,y^*\}} (y-y^*) + f(x^*,y^*).
\end{equation}

which can also be expressed as,

\begin{equation}
P : z = a(x -x^*) +b(y-y^*) + c
\end{equation}

where,

\begin{equation}
a = \dfrac{\partial f}{\partial x} \Big|_{\{x^*,y^*\}},\,\,\, b = \dfrac{\partial f}{\partial y} \Big|_{\{x^*,y^*\}}, \,\,\, c = f(x^*,y^*).
\end{equation}

Then, the multi-point constraint to be imposed to the point $\{x^*,y^*\}$ is given by:

\begin{equation}
u_3 = a u_1 + b u_2 + c - x_3^{i-1},
\end{equation}

Additionally, it is necessary to check that there are no artificial traction forces in the contact interface defined by the MPC. The reaction force $\underline{\mathbf{R}}$ appearing at the contacting nodes, where the MPC have been imposed, has to be checked: the normal contact force should point in the same direction as the normal to the rigid surface:

\begin{equation}\label{eq:contactcondition}
\underline{\mathbf{R}} \cdot \underline{\mathbf{n}} \geq 0.
\end{equation}

\begin{figure*}[h!]
\begin{center}
\includegraphics[width=16.0cm]{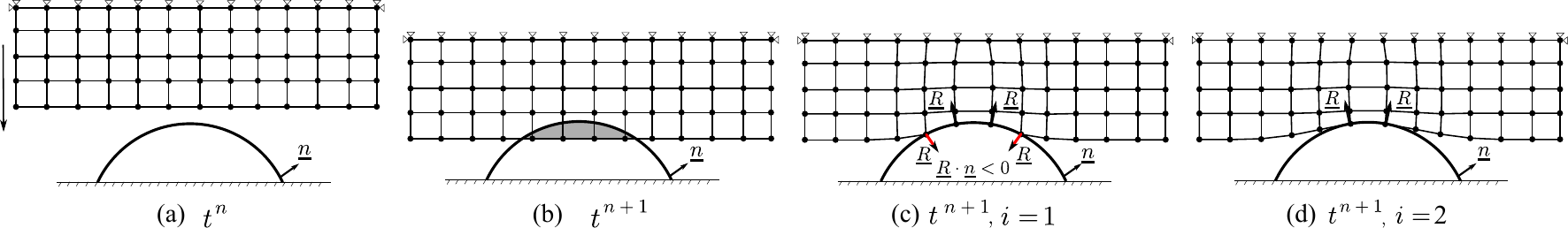} 
\caption{Iterative process for the MPC method. (a) Initial configuration for the two-body contact problem. (b) Four nodes from the deformable body have penetrated to the rigid body. (c) Application of the Dirichlet condition and detection of two adhesion nodes (first iteration). (d) Nodes release and equilibrium are recomputed (second iteration).}
\label{fig:MPC2}
\end{center}
\end{figure*}

Otherwise, the MPC imposed at the contacting node must be removed. Fig.~\ref{fig:MPC2} shows an example of an iterative process for the MPC update process. Once penetrated nodes are detected, MPC boundary conditions are imposed on such nodes. When equilibrium is reached, a search for non-physical adhesion nodes is performed. If adhesion nodes are detected, MPC boundary conditions are released for those nodes and the system is solved again. This procedure is repeated until no adhesion node is found. 

\section{Computational implementation}\label{sec:computimp}

We now present the computational implementation of the parallel algorithm, which solves the unilateral contact problem from a coupled problem point of view. In this approach, each body is treated separately, using one instance of the computational code for each body (multi-code approach). The contact interaction is performed by the transference of Dirichlet boundary conditions at the contact interface. As we tackle a unilateral contact problem (contact between a rigid body and deformable body), the rigid body can be a finite element mesh (current approach) or an analytic expression describing its boundary.

Using a parallel computational code allows us to split each contacting body into several
subdomains. In the implementation presented in this work, the displacements of each body are solved
independently in isolated code instances. Thus, each body can be treated as a standalone problem
with extra boundary conditions due to contact. The mesh is partitioned independently in both code
instances at preprocessing time, not creating any restriction to the mesh partitioner.
In this work we use the graph-based partitioner METIS \cite{Metis}, which maximizes the load balance
while minimizing the sizes of the subdomain interfaces.
The unilateral contact condition is enforced via the transference of information from the rigid to the deformable body. This algorithm can be interpreted as a black-box method, easily adaptable to any finite element computational code. As the solution procedure is performed through the exchange of boundary conditions, this algorithm can be implemented using any parallel computational code since the basis of the contact problem resolution is the exchange of specific boundary information.

To better describe the computational implementation of the contact algorithm, we divide this section into two subsections: (a) contact searching and communication, and (b) contact resolution algorithm. Strong emphasis is dedicated to the latter as it is the main contribution of this work.

\subsection{Contact searching and communication}\label{sec:plepp}

\begin{figure}[h] 
\begin{center}
\includegraphics[width=8.5cm]{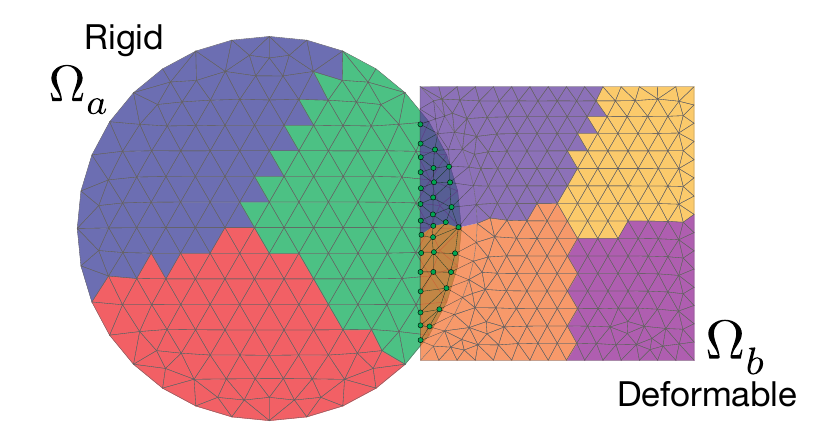} 
\caption{Schematic representation of contact between two bodies partitioned into different subdomains.}
\label{fig:plepp}
\end{center}
\end{figure}

The first step of any contact algorithm is the contact detection. This means to detect interpenetration between the bodies that are prone to contact. In a parallel contact simulation, contact can occur between surfaces owned by different computational nodes (i.e. partitions or subdomains), according to the distribution done by the mesh partitioner. To search for contact, frequent global searches (i.e. unstructured communications) must be done across all the computational nodes. A contact situation in which two partitioned bodies get in contact is depicted in Fig.~\ref{fig:plepp}. Due to the high load of communications that parallel contact detection require, it is a relevant topic in the field of contact mechanics~\cite{Attaway1998}. 

For the localization of contact and communication between subdomains, we use \verb|PLE++|, which is an adaptation of the Parallel Location and Exchange (PLE) library~\cite{Fournier2014}, originally developed by \'Electricit\'e de France (EDF). Due to the black-box characteristics of the algorithm presented in this work, its computational implementation is not restricted to \verb|PLE++| and can be done using any tool capable of detecting overlapping and exchanging information between subdomains. 

\verb|PLE++|~\cite{ZavalaPhd2018} is a parallel 2D and 3D locator and communication tool, used to detect interpenetration and to communicate and transfer information between subdomains. \verb|PLE++| is a C++ environmental library, with the capability of parallel localization of nodes in overlapping subdomains and communication between parallel applications in C, C++, Fortran, or Python. Basically, \verb|PLE++| allows detecting those nodes belonging to $\Omega_b$ which have penetrated into $\Omega_a$, as shown in Fig.~\ref{fig:plepp}. It also performs the inverse operation: identifying the nodes of $\Omega_a$ which have penetrated into $\Omega_b$. As \verb|PLE++| is a parallel tool, the localization is done at the subdomain level. Moreover, \verb|PLE++| allows to communicate and transfer of information between the subdomains involved in the localization process. This is of crucial necessity for solving coupled problems, which are based on the exchange of information between partitions at the boundary interface. 

\begin{figure*}[h]
\begin{center}
\includegraphics[width=14.0cm]{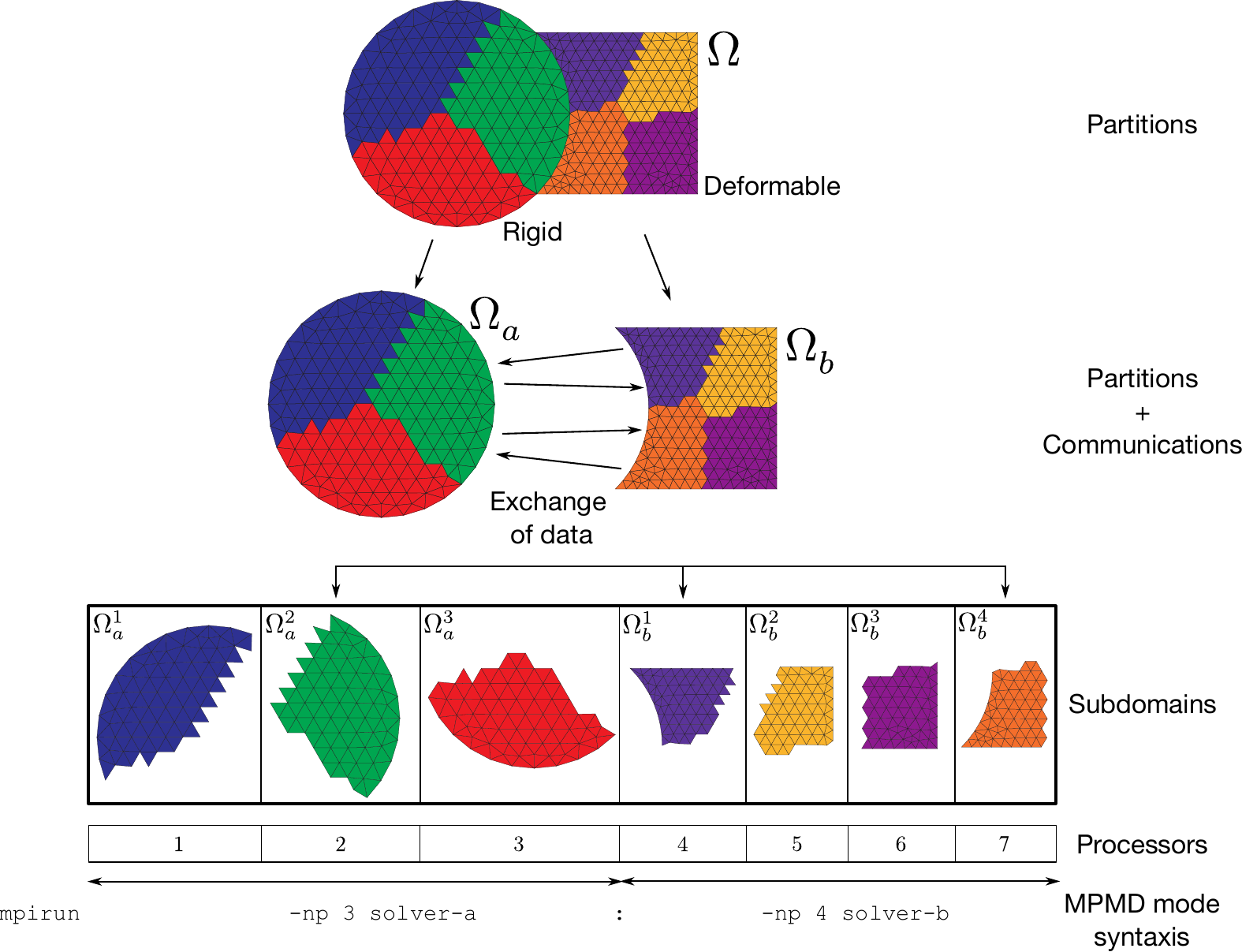} 
\caption{Example of a discretized physical domain formed by two non-conforming meshes with different number of partitions (subdomains).}
\label{fig:partitions}
\end{center}
\end{figure*}

In Fig.~\ref{fig:partitions} we show a discretized physical domain formed by two non-conforming partitions $\Omega_a$ and $\Omega_b$. A total of seven processors are used to solve the system, distributed as follows: three processors are assigned to $\Omega_a$ and four to $\Omega_b$. Some boundary nodes of the subdomain $\Omega_a^2$ (processor 2) are in contact with the interface boundary of subdomains $\Omega_b^1$ (processor 4) and $\Omega_b^4$ (processor 7). Thus, only processors 2, 4, and 7 are involved in the coupling between partitions $\Omega_a$ and $\Omega_b$. In order to establish the connections needed for the resolution of the contact problem, each processor must know exactly which of its nodes (if any) are contained by any other subdomain. On the other hand, if a given processor contains external nodes (i.e. nodes owned by other processors), it must know the number and coordinates of those nodes. Also, that processor must know which of its elements contain each of the external nodes and the processor to which those nodes belong. To summarize how \verb|PLE++| does the contact searching and communication its main algorithms are described below. The reader is referred to~\cite{ZavalaPhd2018} for more details.

\subsubsection{Searching and localization}\label{sec:searching}

\begin{algorithm}[h!]
\caption{Search/localization}
\label{alg:global_search}
\begin{algorithmic}[1]
\State $Q_{a}^{k}  = GetBox(\boldsymbol \Omega^{k}_{a})$ \label{lin:gs1}
\For{$l =\{ 1,2,..., p_{b} \}$ } \label{lin:gsfor}
  \State $Q^{l}_{b}  = SendRecv_l( Q_{a}^{k}  )$ \label{lin:gs2}
  \If{$Q_{ab}^{kl} = Q^{k}_{a} \cap Q^{l}_{b} \neq \emptyset$} \label{lin:gs3}
    \State $\textbf{Id}_{a}^{k}   = WithinBox( \textbf{R}_{a}^{k}  , Q^{k}_{a} \cap Q^{l}_{b} )$  \label{lin:gs4}
    \State $\textbf{r}_{a}^{k} = \textbf{R}_{a}^{k}( \textbf{Id}_{a}^{k} )$ \label{lin:gs5}
    \State $\textbf{Id}_{a_c}^{k}   = WithinBox(\boldsymbol \Omega_{a}^{k}  , Q^{k}_{a} \cap Q^{l}_{b} )$ \label{lin:gs51}
    \State $\boldsymbol \omega^{k}_{a} = \boldsymbol \Omega^{k}_{a}( \textbf{Id}_{a_c}^{k} )$ \label{lin:gs52}
    \State $\textbf{r}_{b}^{l}  = SendRecv_l( \textbf{r}_{a}^{k} )$ \label{lin:gs6}
    \State $\textbf{Id}_{a_c}^{ \textbf{r}_b}   =  LocateInCells(  \textbf{r}_{b}^{l} , \boldsymbol \omega^{k}_{a} )$ \label{lin:gs7}
    \State $\textbf{Id}_{a_c}^{ \textbf{r}_a }  = SendRecv_l(  \textbf{Id}_{a_c}^{ \textbf{r}_b} )$ \label{lin:gs8}
    \State $\textbf{ID}_{a}^{k} [l] = ( \textbf{Id}_{a}^{k},  \textbf{Id}_{a_c}^{ \textbf{r}_{a} } )$ \label{lin:gs9}
  \EndIf 
\EndFor
\end{algorithmic}
\end{algorithm}

In Algorithm~\ref{alg:global_search}, we show how \verb|PLE++| performs the first step named as searching and localization of the contact nodes. For the sake of clarity, we are going to suppose the case where a local processor, $k$ is that belonging to $\Omega_a$ while the remote ones, $l$, are those from $\Omega_b$. First of all, a global search is performed creating a global bounding box for each of the subdomains of bodies $\Omega_a$ and $\Omega_b$. Let's assume that this algorithm is executed in a local processor $k$. The first task of the algorithm is to share with all remote processors (line 2) the geometric definition of each local bounding box (line 3). After this information is shared with all subdomains through the \verb|SendRecv| instruction, each local processor compares its bounding box with all bounding boxes received from the remote processors. When there is an overlap between subdomains from different bodies (line 4), the next step is to store the list of node identifiers $\textbf{Id}_{a}^{k}$, where $\textbf{R}_{a}^{k}$ is a list of local nodes which lie inside the overlapping region $Q_{ab}^{kl}$. The coordinates list $\textbf{r}_{a}^{k}$ (line 6) is obtained through the array of the local nodes. Additionally, it is important to know the information about the elements and their nodal connectivity. The element identifier $\textbf{Id}_{a_c}^{k}$ is an array of local element numbering from $\boldsymbol {\Omega}_{a}^{k}$, which allows to store the element connectivities $\boldsymbol \omega^{k}_{a}$ (line~\ref{lin:gs52}). After that, the subset of local nodes $\textbf{r}_{a}^{k}$ is shared with all remote processors which fulfill the condition $Q_{a}^{k} \cap  Q_{b}^{l} \neq \emptyset$ (line~\ref{lin:gs6}). After this information is collected, a local search to identify which remote nodes lie inside the local elements (line~\ref{lin:gs7}) is conducted. This is performed using an octree search and linear interpolation using barycentric coordinates strategies, see~\cite{ZavalaPhd2018}. Finally, the last step of the algorithm is to share the identifiers $\textbf{Id}_{a_c}^{ \textbf{r}_b}$ to the remote processor with the \verb|SendRecv| instruction (line~\ref{lin:gs8}) and to store the node identifiers $\textbf{Id}_{a}^{k}$ and the element identifiers $\textbf{Id}_{a_c}^{ \textbf{r}_a}$ in the assembly array $\textbf{ID}_{a}^{k} [l]$ (line~\ref{lin:gs9}). This array serves to conduct the exchange of information for the coupling between bodies. 

\subsubsection{Data exchange}\label{sec:exchange}

Algorithm~\ref{alg:data_exchange} describes the data exchange strategy. The communication between one of the subdomains $\Omega_a^{k}$ and another from $\Omega_b^{l}$ (and vice versa) only occurs when $Q_{a}^{k} \cap  Q_{b}^{l} \neq \emptyset$ condition is fulfilled (line~\ref{lin:de1}). For instance, in the example shown in Fig.~\ref{fig:partitions}, the exchange would be between $\Omega_a^{1}$ and the two subdomains $\Omega_b^{1}$ and $\Omega_b^{4}$. The exchange step consists of doing first the interpolation of the local properties $\textbf{v}_{a}^{k}$ to the remote nodes ($\textbf{r}_{b}^l$) with the aim of getting the interpolated values $\textbf{v}_{b}^{l}$ at $\Omega_b^{l}$ (line~\ref{lin:de2}). This is done locally in each processor and then the values are communicated using the Message Passing Interface (MPI). In the case of unilateral contact problems, the exchange of data related to the contact force can be straightforward without any interpolation. The rigid body only receives the force, which is the sum of the nodal forces of the deformable body at the contact region. On the contrary, the contact between deformable bodies requires conservative interpolation methods for the exchange of data between bodies, see~\cite{ZavalaPhd2018} and~\cite{Rivero2018PhD}.

\begin{algorithm}[h]
\caption{Data exchange}
\label{alg:data_exchange}
\begin{algorithmic}[1]
\For{$l =\{ 1,2,..., p_{b} \}$ } \label{lin:de0}
\If{$Q_{ab}^{kl} = Q^{k}_{a} \cap Q^{l}_{b} \neq \emptyset$} \label{lin:de1}
      \State Interpolation$_{a}$($\textbf{v}_{a}^{k},\textbf{r}_{b}^l$) $\rightarrow \textbf{v}_{b}^{l}$ \label{lin:de2}
      \State Communication$_{ab}$($\textbf{v}_{b}^{l}$) \label{lin:de3}
  \EndIf
\EndFor 
\end{algorithmic}
\end{algorithm}

\subsection{Contact algorithm}\label{sec:contalago}

One of the most important advantages of the proposed contact algorithm is its flexibility to solve each body separately depending on the complexity of the problem. In the present paper, the resolution of the rigid body is performed using a Runge-Kutta 4$^{th}$ order scheme \cite{Griffiths2006} while the deformable body is solved using a Newmark $\beta$ implicit scheme \cite{Belytschko2014}, but any solution scheme suitable for each code instance can also be used. The solution scheme used to solve the deformable body is designed to be used in implicit solvers.

\begin{figure*}[h]
\begin{center}
\includegraphics[width=16cm]{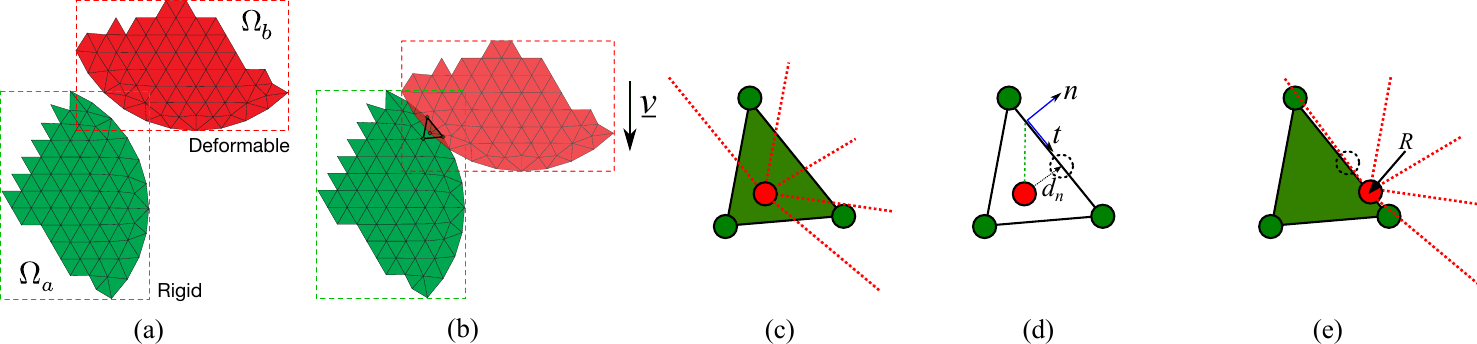} 
\caption{Workflow of the parallel contact algorithm (contact searching and resolution). (a) Interaction. (b) Overlapping. (c) Contact searching. (d) Projection and Dirichlet condition. (e) Equilibrium.}
\label{fig:synch}
\end{center}
\end{figure*}

\begin{figure}[h]
\begin{center}
\includegraphics[width=7.0cm]{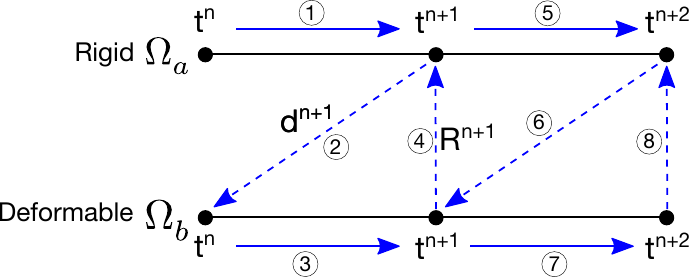} 
\caption{Multicode Gauss-Seidel coupling algorithm for the parallel contact between rigid and deformable bodies.}
\label{fig:gausseidel}
\end{center}
\end{figure}

The workflow of the parallel contact algorithm is shown in Fig.~\ref{fig:synch} and the algorithm as a coupling scheme is shown in Fig.~\ref{fig:gausseidel}. The execution is done in a staggered way, so the first execution is the rigid body and then follows the deformable body, see  Fig.~\ref{fig:gausseidel}. This solution process can be interpreted as a Gauss-Seidel coupling scheme with weak coupling. At each time step, contact detection is done after the displacements update of the rigid body but before the resolution loop of the deformable body. In other words, we use the updated mesh of the rigid body at time $t^{n+1}$ as the base mesh for the localization of penetrated nodes of the previous time step ($t^{n}$) mesh of the deformable body. At this instant, both instances synchronize and the localization is executed. This procedure is repeated at each time step and the contact algorithm is triggered only when at least one boundary node of the deformable body has penetrated inside the rigid body. When this happens, the rigid body computes and sends to the deformable body all the information required for the enforcement of the MPC boundary conditions, which are used as additional constraints to find the equilibrium configuration of the deformable body. In the following subsections, we explain in further detail this procedure.

\subsubsection{Rigid body}

\begin{figure}[h!] 
\begin{center}
\includegraphics[width=8.5cm]{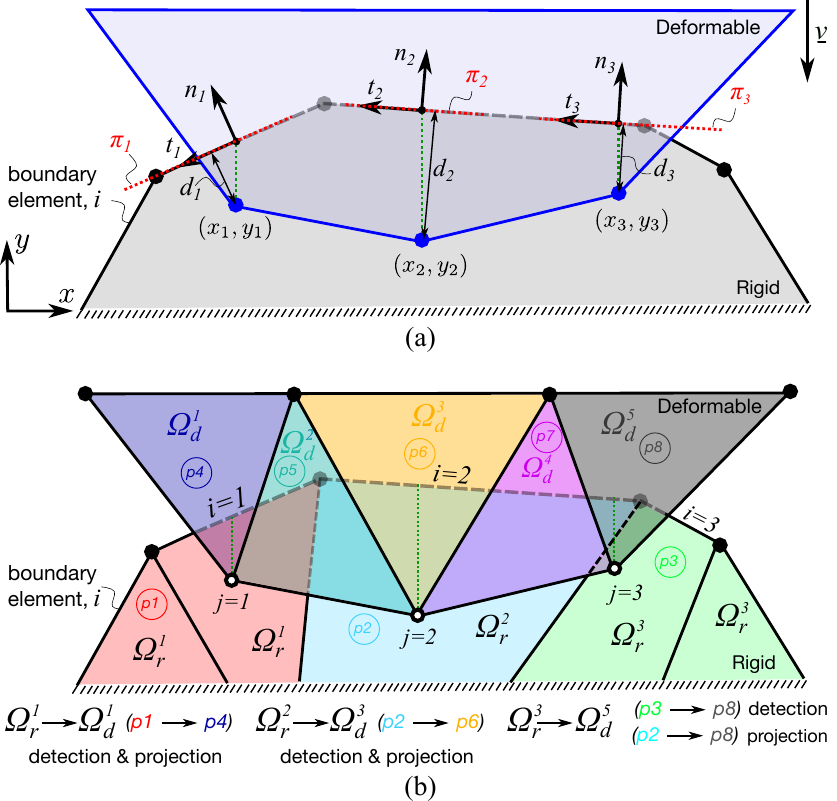} 
\caption{(a) Projection operation done by the processors which belong to the rigid body. (b) Transference  of information from the rigid body to the deformable body. The reader is referred to the web version of this paper for the color representation of this figure.}
\label{fig:projec_and_transfer}
\end{center}
\end{figure}

At this stage of the contact algorithm, the aim is to compute the local coordinate system for each contacting node and its distance to the rigid body boundary, which will be later used as an MPC in the deformable body, as shown in Fig.~\ref{fig:projec_and_transfer}a. When contact is detected, the projection algorithm (Algorithm~\ref{alg:rigid}) is executed concurrently (i.e. in parallel) in all the processors which belong to the rigid body partition. The projection direction $v$ must be given as an input by the user and it is used for determining the boundary element where the local system has to be created. This algorithm is sufficiently general in order to consider the case where the projection lies in a boundary segment of a different processor than the one which owns the detected node (see Fig.~\ref{fig:projec_and_transfer}b, node 3). The strategy followed in this work is to communicate to all the processors of the rigid body partition the information of the complete contact boundary definition (lines~\ref{lin:rig1} to~\ref{lin:rig4}). After line~\ref{lin:rig4} is executed, each processor of the rigid body partition has a local copy of the coordinates of the boundary nodes and the boundary element connectivities of the complete contact boundary of the rigid body. For each of the {\it npoin\_loc\_b} nodes of the deformable body that have penetrated the rigid body (line~\ref{lin:rig5}), the algorithm loops over each boundary element {\it nboun\_loc\_a} of the contact boundary of the rigid body (line~\ref{lin:rig6}) and builds a plane $\pi_i$ (or a line, in a 2D case) using the geometrical information of the boundary element (line~\ref{lin:rig7}). Then, the algorithm projects the detected node $j$ to the plane $\pi_i$ in the direction $v$ (line~\ref{lin:rig8}). Afterwards, the algorithm checks if the projection of the node $j$ to plane $\pi_i$ lies inside the boundary element $i$ (line~\ref{lin:rig9}). If this is true, the algorithm returns the normal vector, $n_{j}$, tangent vector, $t_{j}$, and normal distance, $d_{j}$ from node $j$ (lines~\ref{lin:rig10} to~\ref{lin:rig11}). This procedure is schematically represented in Fig.~\ref{fig:projec_and_transfer}. The output of this computation is a normal-tangent orthonormal basis for each of the penetrated nodes. This orthonomal basis is created on the boundary element defined by the projection of the contact node, see Fig.~\ref{fig:projec_and_transfer}a. The normal distance, $d_{j}$, which corresponds to the distance from the point to the normal tangent line, is also computed. 

\begin{algorithm}[h!]
\caption{Rigid body algorithm (Code 1)}
\label{alg:rigid}
\begin{algorithmic}[1]
\Require $v$, \emph{npoin\_loc\_b}
\Return $n$, $t$, $d$    
\State Communicate local number of rigid body boundary elements                            \label{lin:rig1}
\State Get total number of rigid body boundary elements: \emph{nboun\_loc\_a}          \label{lin:rig2}
\State Communicate local boundary nodes coordinates                             \label{lin:rig3}
\State Get all boundary nodes coordinates : \emph{bocod\_loc\_a}                 \label{lin:rig4}
\For{$j = 1,npoin\_loc\_b$}                                                               \label{lin:rig5}
  \For{$i = 1,nboun\_loc\_a$}                                                 \label{lin:rig6}
    \State Create plane $\pi_i: ax + by + cz +d$                                   \label{lin:rig7} 
    \State Projection node $j$ to plane $\pi_i$ using $v$                   \label{lin:rig8}
    \If{Projection is inside boundary element $i$}                              \label{lin:rig9}
      \State Compute local basis: $n_{j}$, $t_{j}$ \label{lin:rig10}
      \State Compute normal distance: $d_{j}$      \label{lin:rig11}
    \EndIf
  \EndFor
\EndFor                     
\end{algorithmic}
\end{algorithm}

After the rigid body's code instance has computed all the information required for each contacting node, the next step is to send this data (exchange) to the deformable body, as shown Fig.~\ref{fig:gausseidel} (step 2). This exchange of information is done via the \verb|PLE++| tool, described in Sec.~\ref{sec:plepp}.
Each processor of the rigid body that has detected a contacting node will send the orthogonal basis and normal distance to the processor of the deformable body which owns that node. Fig.~\ref{fig:projec_and_transfer}b shows an example where the rigid body is divided in 3 subdomains while the deformable body is divided in 5 subdomains. Node 1 ($j=1$), which belongs to the deformable body's contact surface is detected by processor 1 (subdomain 1) of the rigid body. Once the orthogonal basis and normal distance are computed, this information is sent to processor 4 (subdomain 1) of the deformable body, as this is the processor/subdomain which owns that node. A similar procedure is performed for node 3 ($j=3$), where the projection of this node lies on a boundary which belongs to processor 2 but it is detected by processor 3. In this case, the projection is transferred to the appropriate boundary (subdomain 2) as each of the subdomains of the rigid body know the complete contact boundary information of the whole domain.

\subsubsection{Deformable body}

\begin{figure}[h] 
\begin{center}
\includegraphics[width=8.0cm]{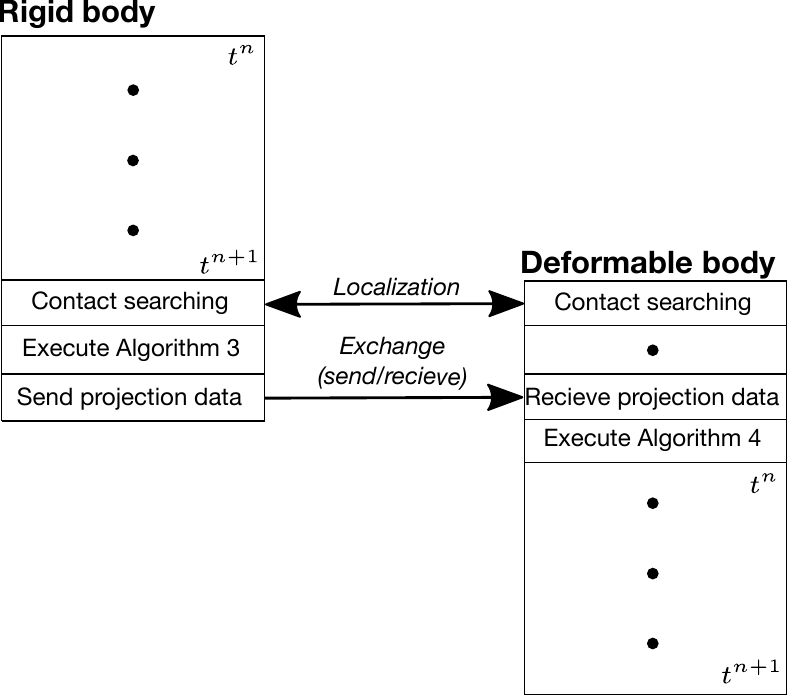} 
\caption{Staggered execution of rigid and deformable body code instances for the resolution of contact at $t^{n+1}$.}
\label{fig:staggered_execution}
\end{center}
\end{figure}

The exchange of the projection data between the rigid and deformable body is produced after the end of the execution of Algorithm~\ref{alg:rigid}. This means that after the contact detection, the code instance in charge of the solution of the deformable body is on hold waiting for the rigid body instance to finish the execution of Algorithm~\ref{alg:rigid}. Then, the execution of the code instance in charge of the rigid body stops and the Algorithm~\ref{alg:deformable}, which is executed concurrently by each processor that belongs to the code instance in charge of the deformable body, starts. This staggered execution is shown in Fig.~\ref{fig:staggered_execution}.

\begin{algorithm}[h]
\caption{Deformable body algorithm}
\label{alg:deformable}
\begin{algorithmic}[1]
\Require \emph{npoin\_loc\_b}, $n$, $t$, $d$
\State Initialize all nodes: \emph{tag\_contact\_j} $\gets$ 0
  \For{$j = 1,npoin\_loc\_b$}                                                    \label{lin:defo2}
    \State \emph{tag\_contact$_j$} = 1                                              \label{lin:defo3}
    \State Construct rotation matrix $T_{j}$ with $n_j$, $t_{j}$  \label{lin:defo4}
    \State Set Dirichlet displacement, $d_j$          \label{lin:defo5}
  \EndFor
\end{algorithmic}
\end{algorithm}

Each of these processors loop over the {\it npoin\_loc\_b} nodes (line~\ref{lin:defo2}). These nodes, which belong to the boundary of the deformable body, are marked as contact nodes (line~\ref{lin:defo3}). Next, a rotation matrix is built for each of these nodes using the information received by the rigid body (line~\ref{lin:defo4}). Finally, a prescribed displacement (Dirichlet condition) is applied to each of the {\it npoin\_loc\_b} nodes using the normal distance $d_j$ to the tangent plane (line~\ref{lin:defo5}). Both, the rotation matrix and the prescribed displacement are used for the enforcement of the MPCs.

The effect of the MPC for each contacting node is that they restrict their movement along the tangential plane located at the projected point in the contact surface of the rigid body, see Fig.~\ref{fig:projec_and_transfer}a. To avoid the introduction of additional degrees of freedom in order to solve this kind of constraint, we rotate locally the coordinate system of each node $j$ using the matrix $T_j$ and enforce a homogeneous Dirichlet boundary condition along the normal direction. Once we get the equilibrium solution of the deformable body, we rotate back the local coordinate system of all the contacting nodes, to transform the solution to the global $x-y$ reference frame.

For a general 3D case, the rotation matrix $T_j$ for the node $j$ is constructed as follows: 

\begin{equation}\label{ec:rotation_matrix}
T_j = \left[ {\begin{array}{ccc}
n_x & t_{1,x} & t_{2,x} \\
n_y & t_{1,y} & t_{2,y} \\
n_z & t_{1,z} & t_{2,z} 
\end{array}} \right],
\end{equation}

where $\mathbf{n}$ is the normal vector and $\mathbf{t}_1$ and $\mathbf{t}_2$ the tangential vectors associated to node $j$. In a 2D problem, only one tangent vector is needed. 

The algorithm which controls the system resolution and the nodes' release in a parallel execution is shown in Algorithm~\ref{alg:reso}. This algorithm is executed concurrently in each processor of the deformable body partition at each time step.

\begin{algorithm}[h!]
\caption{System resolution algorithm for an implicit time integration scheme}
\label{alg:reso}
\begin{algorithmic}[1]
  \State Update boundary conditions                                     \label{lin:reso1}
  \State Assembly of right-hand-side and system matrix     \label{lin:reso2}
  \State Set initial guess for inner iterations                           \label{lin:reso3}
  \State \emph{not\_converged} $\gets$ 1
  \While{\emph{not\_converged} == 1}                                      \label{lin:reso5}
    \State call Newmark-$\beta$ implicit scheme                   \label{lin:reso6}
    \State Check convergence Newton iterations                    \label{lin:reso7}    
    \If{\emph{not\_converged} == 0}                                            \label{lin:reso8}
      \State Compute reactions $\underline{\mathbf{R}}$ on each contact node $i$ \label{lin:reso9}
      \If{\emph{nodes\_to\_release}}                                 \label{lin:reso10}
        \State Release nodes                                         \label{lin:reso11}
        \State \emph{not\_converged} = 1           \label{lin:reso12}
      \EndIf                                                                    \label{lin:reso13}
    \EndIf                                                                      \label{lin:reso14}
    \State call \verb|MPI_SUM|( \emph{not\_converged} )         \label{lin:reso15}
    \If{\emph{not\_converged} $>=$ 1}                                       \label{lin:reso16}
      \State \emph{not\_converged} = 1                             \label{lin:reso17}
    \EndIf                                                                     \label{lin:reso18}
  \EndWhile                                                               \label{lin:reso19}
\end{algorithmic}
\end{algorithm}

After the contact constraints enforcement, the resultant linear system of equations for the deformable body is solved. The result of this procedure is the equilibrium configuration of the deformable body given the MPC imposed by contact (lines~\ref{lin:reso1} to~\ref{lin:reso7} of Algorithm~\ref{alg:reso}). Once the convergence is achieved, reactions are calculated in each contact node (line~\ref{lin:reso9}). The next step is to check if any of the contact nodes is in artificial adhesion (line~\ref{lin:reso10}). If this is the case, the MPC of the node is released (line~\ref{lin:reso11}) and the system is solved again. This procedure is repeated until all contact nodes in artificial adhesion are released. 

In general, the nodes which are subjected to artificial adhesion are owned by subset of all the processors which own the totality of the contact nodes. As Algorithm~\ref{alg:reso} is executed concurrently by all processors, this implies that the processors which own the adhered nodes will continue the execution of the algorithm while the rest will exit the loop as they converged. As it is necessary to keep all processors executing Algorithm~\ref{alg:reso} once adhesion nodes are detected, it is crucial to synchronize the execution of all processors. This is done by communicating to all the processors the convergence flag (lines~\ref{lin:reso15} to~\ref{lin:reso18}). By doing this, it is assured that all processors exit the loop simultaneously when no adhesion node is longer detected.

\section{Benchmark tests}\label{sec:benchmark}

In this section we verify and validate the parallel contact algorithm in 2D and 3D examples. The algorithm described in the present paper is implemented within an environment designed for heterogeneous problems in computational mechanics, which is the multiphysics, multiscale and massively parallel code \verb|Alya| \cite{Vazquez2016a}. Taking profit of the flexibility and generality of the algorithm presented in this work we use a multicode scheme. Under this scheme, the displacement field of each body (rigid and deformable) is solved using different instances of the BSC's in-house Finite Element Code, \verb|Alya|, while the contact detection and exchange of contact information between instances is done by \verb|PLE++| \cite{ZavalaPhd2018}. It is worth highlighting that in all the benchmarks presented in this section, both rigid and deformable bodies are discretized with a finite element mesh to assess the geometrical localization algorithm. This methodology is flexible enough to deal with any kind of geometry and it enables the extension of the current unilateral contact algorithm to deformable-to-deformable contact problems. The Implicit Newmark-Beta time integration scheme \cite{Belytschko2014} using $\beta=0.65$, $\gamma=0.9$ and $\alpha=0$ is used for the non-linear iterations of the deformable body and an iterative Conjugate-Gradient solver with diagonal preconditioning is used for the resolution of the algebraic system. With respect to the rigid body, a 4$^{th}$ order Runge-Kutta scheme \cite{Griffiths2006} is used in all cases. All the simulations are conducted in {\it MareNostrum4} supercomputer. This cluster has 3456 nodes, each of them with 48 cores Intel Xeon Platinum $@$ 2.1 [GHz], giving a total core count of \num{165888} cores.
 
\subsection{Hertz contact problem}

\begin{figure}[H]
\begin{center}
\includegraphics[width=4cm]{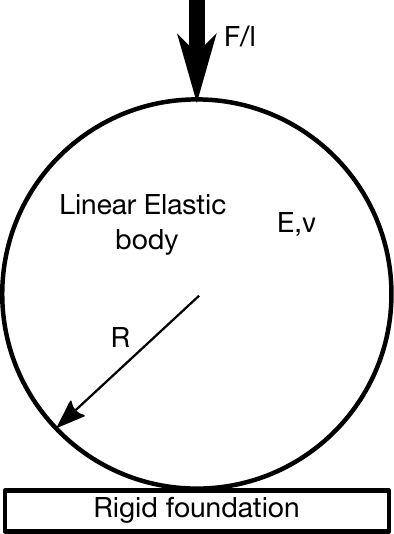} 
\caption{Physical model of the Hertz contact problem}
\label{fig:hertz-physical-problem}
\end{center}
\end{figure}

The Hertz contact problem is a classical example for verifying the physics of the proposed contact algorithm. This problem consists of a circular and infinitely long cylinder of radius $R$ and length $l$ resting on a flat rigid foundation. The cylinder is subjected to an uniform load along its top $F/l$ and it is constructed of an homogeneous, isotropic, elastic material with Young's modulus $E$ and Poisson's ratio $\nu$, see Fig.~\ref{fig:hertz-physical-problem}. Taking that $l >> R$, we can assume a problem of plane strain. No friction is assumed to exist on the contact surface. 

The Hertz solution (see~\cite{Kikuchi1988}) yields a contact pressure distribution of:

\begin{equation}\label{eq:hertz_pressure}
P = \frac{2 F}{\pi b^2 l}\sqrt{(b^2 - x^2)},
\end{equation}

where $F/l$ is the load per unit length and $b$ is the half-width of contact surface defined by:

\begin{equation}\label{eq:hertz_width}
b = 2\sqrt{\frac{F R (1-\nu^2)}{\pi l E}}.
\end{equation}

\begin{figure*}[h]
\begin{center}
\includegraphics[width=12cm]{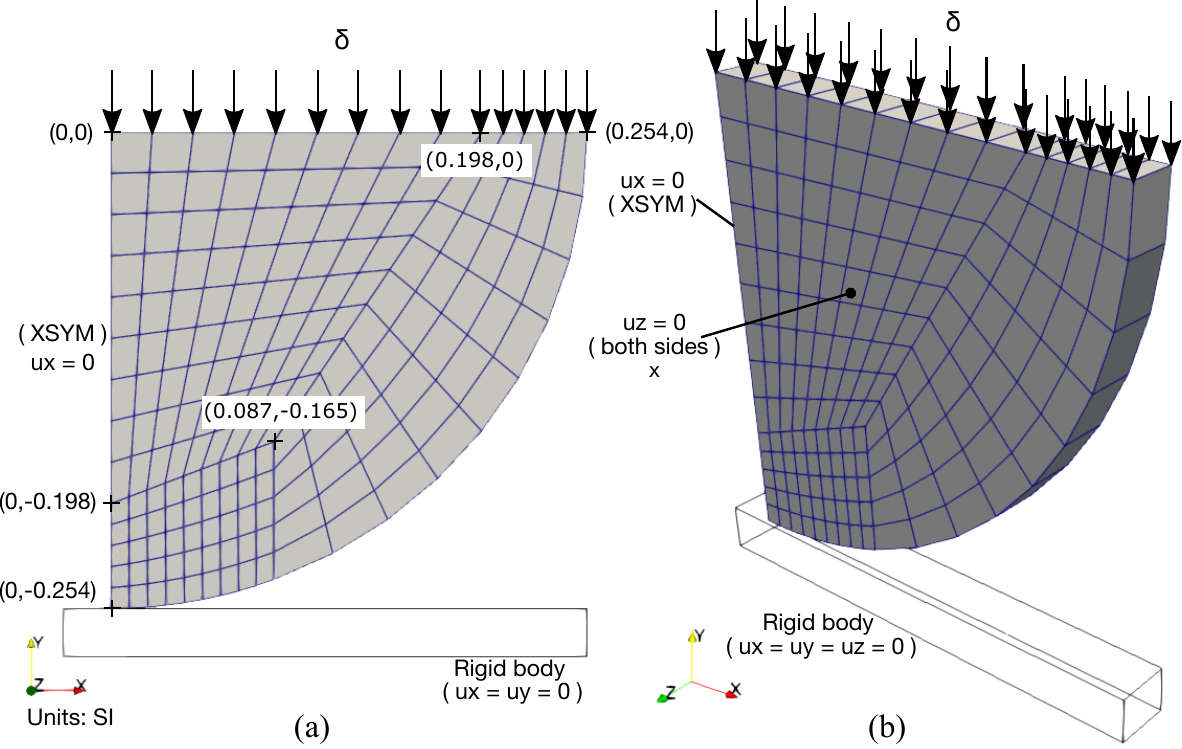} 
\caption{Numerical models of the Hertz contact problem (a) 2-dimensions and (b) 3-dimensions.}
\label{fig:numerical-model}
\end{center}
\end{figure*}

We solve an equivalent problem in 2D and 3D, as depicted in Fig.~\ref{fig:numerical-model}. In both models, we apply a prescribed displacement at the topmost edge (2D model) or face (3D model). Taking profit the symmetry of the problem only one-quarter of the cylinder is modeled. It is worth mentioning that the 3D case, a cylinder of unit thickness is modeled. In order to impose the plane strain condition, the out-of-plane displacements are fixed on the two exterior faces, see Fig.~\ref{fig:numerical-model}b. The 2D model is meshed with quadrilateral plane strain elements, while the 3D case is meshed with 8-node linear solid elements using full integration.

In this example, the cylinder has a radius of \SI{0.254}{\m}, Young's modulus of \SI{206}{\giga\pascal} and Poisson's ratio of \SI{0.3}{}. The vertical displacement imposed is \textbf{\SI{10.16e-3}{\m}}. The total time of the simulation is set to \SI{1}{\second} with a prescribed time step of 0.1. The maximum contact pressure obtained for each numerical model is $P_{max}=\,$\SI{18.427}{\giga\pascal} which corresponds to a distance of ($x = 0$). By combining Eqs.~\eqref{eq:hertz_pressure} and~\eqref{eq:hertz_width} we can isolate the variable $F/l$ in order to compute $b$, which results in $b=\,$\textbf{\SI{41.18e-3}{\m}}.

\begin{figure}[h!]
\begin{center}
\includegraphics[width=8.0cm]{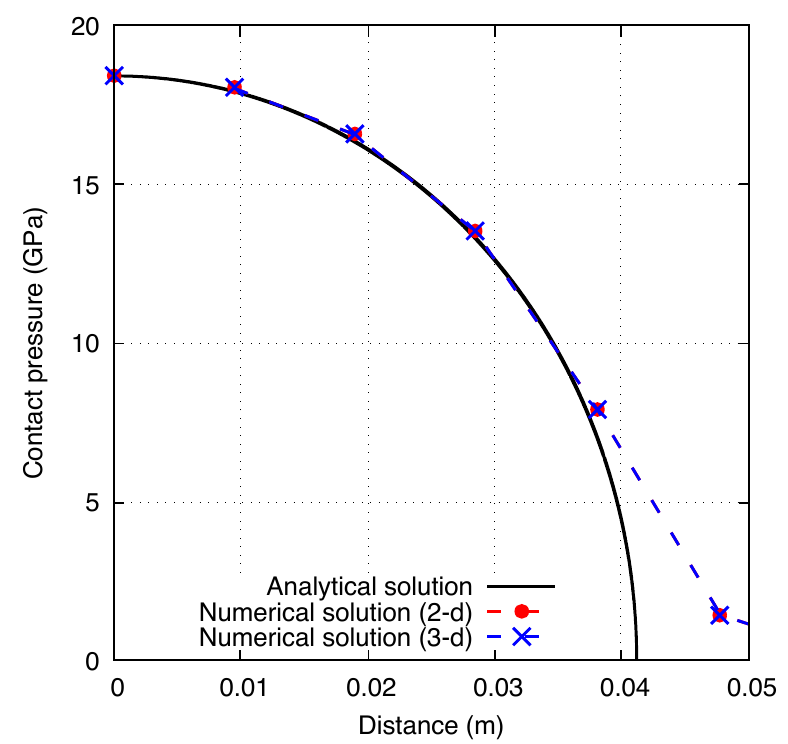} 
\caption{Contact pressure versus position for the Hertz contact problem.}
\label{fig:hertz-verification}
\end{center}
\end{figure}

A comparison of the computed contact pressure with the Hertz solution (Eq.~\eqref{eq:hertz_pressure}) is given in Fig.~\ref{fig:hertz-verification}. We can observe a good agreement between analytical and numerical solutions for both dimensions. When the distance is higher than \SI{0.04}{\m}, the contact pressure differs from the theoretical solution due to the coarse discretization in this rounded region. On the contrary, the contact pressure is accurately predicted at the center region where the mesh is refined. Thus, it is advisable to have a dense mesh or even quadratic elements when dealing with rounded shapes.

\subsection{3D indentation test}

\begin{figure*}[h!]
\begin{center}
\includegraphics[width=12cm]{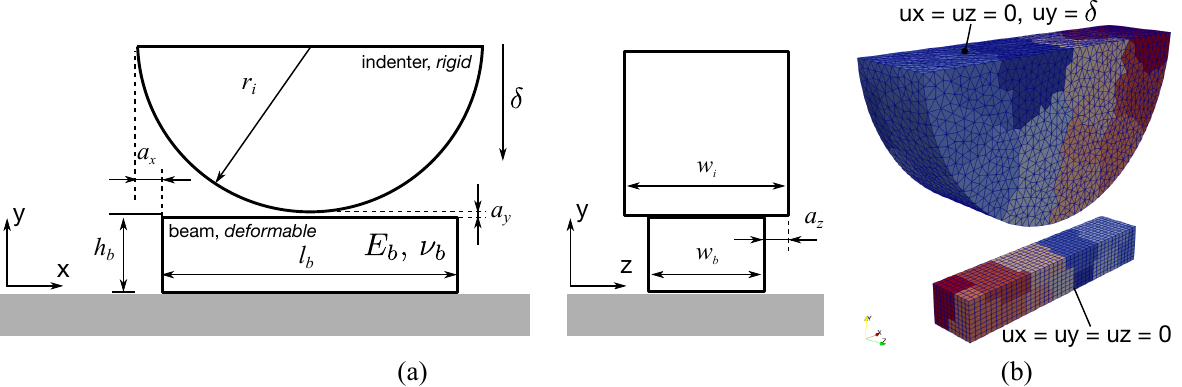} 
\caption{(a) Physical model of the 3-d indentation test. (b) Domain decomposition of the mesh and boundary conditions \cite{Rivero2018PhD}.}
\label{fig:indentation3d-model}
\end{center}
\end{figure*}

This example solves a 3D indentation test, which consists of a rounded-head rigid indenter and a deformable beam, see M. Rivero \cite{Rivero2018PhD} for more details. The physical model for this problem is shown in Fig.~\ref{fig:indentation3d-model}a. The dimensions of the rigid body (indenter) are: $r_i=\,$\SI{1}{\m} and $w_i=\,$\SI{0.5}{\m}, while the deformable body (beam) are $h_b=\,$\SI{0.25}{\m}, $l_b=\,$\SI{1.5}{\m} and $w_b=\,$\SI{0.3}{\m}. The relative position of the indenter with respect to the beam is given by $a_x=\,$\SI{0.25}{\m} and $a_z=\,$\SI{0.1}{\m} and $a_y=\,$\SI{0.01}{\m} (gap). The beam is modeled with an hyperelastic Neo-Hookean formulation \cite{AnsysMechanical} and finite strains, with material properties $E_b=\,$\textbf{\SI{6.896e+8}{\pascal}} (Young modulus) and $\nu_{b}=\,$\SI{0.32}{} (Poisson ratio). The prescribed displacement by the indenter along the vertical axis is $\delta =\,$\SI{0.11}{\m}, while the bottom part of the beam is fixed in all directions. The beam is meshed with 8-node linear solid elements and the indenter is meshed with 4-node linear tetrahedrons, both element types with full integration. The mesh of the model contains \num{224640} elements for the beam and \num{15960} for the indenter. A non-linear static analysis is performed with a total time of the simulation of $\SI{1}{\s}$ and a fixed time step of \num{0.025}.

To test the parallel behavior of the contact algorithm we solve this problem using 12 subdomains for the beam and 24 for the indenter, see Fig.~\ref{fig:indentation3d-model}b. It is important to remark that the contact zone shares several subdomains, so the localization of nodes and their communication for data transfer between instances and partitions is done dynamically.

\begin{figure}[h]
\begin{center}
\includegraphics[width=7cm]{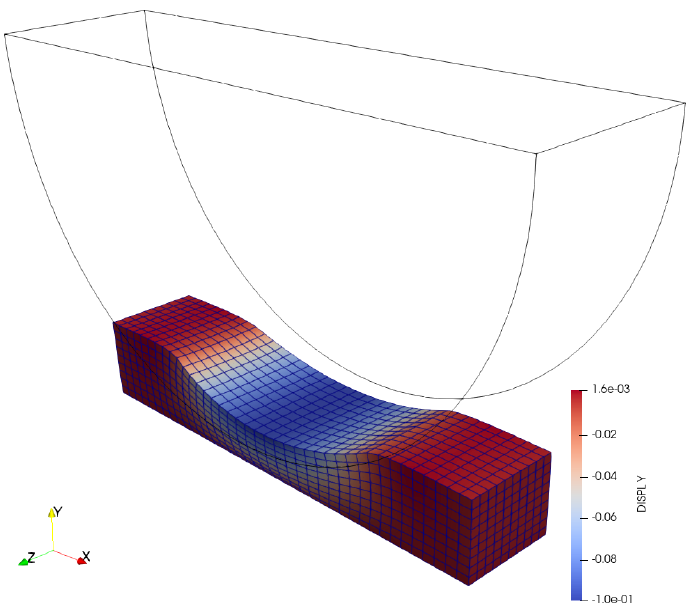} 
\caption{Beam's displacement in y-direction due to an indenter's displacement of \SI{0.11}{\m} and assuming an initial gap of \SI{0.01}{\m}.}
\label{fig:deformed-indentation3d}
\end{center}
\end{figure}

\begin{figure}[h]
\begin{center}
\includegraphics[width=7cm]{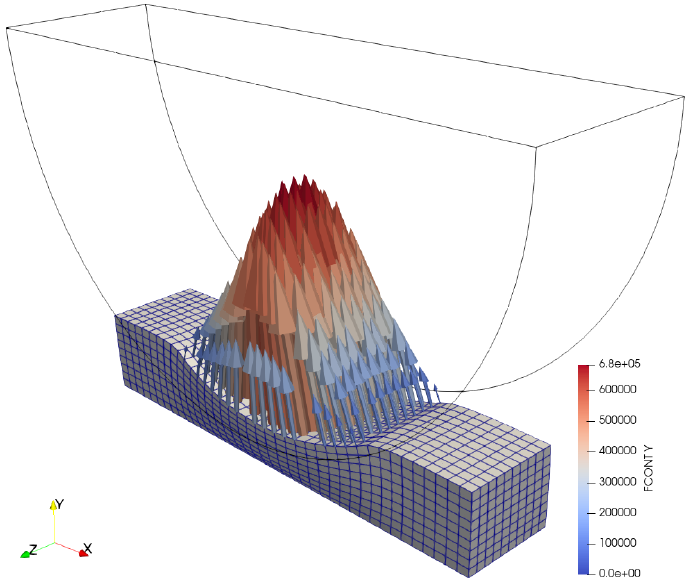} 
\caption{Contact forces resulting from an indenter's displacement of \SI{0.11}{\m} and assuming an initial gap of \SI{0.01}{\m} (scaled factor for vectors is \textbf{\SI{1.e-7}{}}).}
\label{fig:forces-indentation3d}
\end{center}
\end{figure}

The deformed shape of the beam and the forces originated at the contact zone are respectively shown in Fig.~\ref{fig:deformed-indentation3d} and Fig.~\ref{fig:forces-indentation3d}.

\begin{figure*}[h]
\begin{center}
\includegraphics[width=16cm]{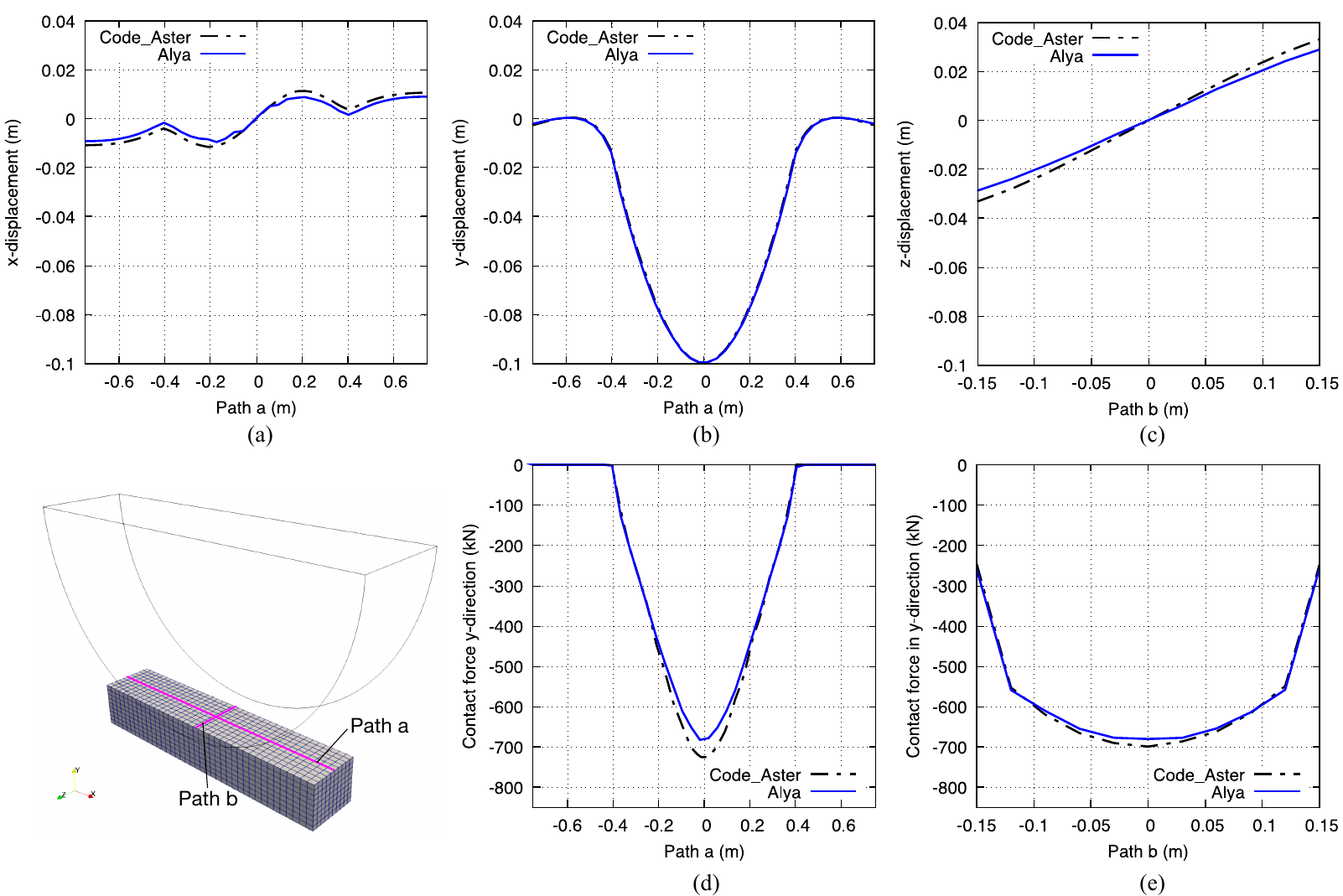} 
\caption{Displacements and contact forces for path $a$ and $b$ \cite{Rivero2018PhD}. (a) Vertical displacement. (b) Tangential displacement path $a$. (c) Tangential displacement path $b$. (d) Contact force at path $a$. (e) Contact force at path $b$.}
\label{fig:resultspaths}
\end{center}
\end{figure*}

In this example the numerical predictions are compared against the solution obtained by the FE code \verb|Code_Aster| \cite{AsterFE}. The different displacements and forces at the contact region are shown in Fig.~\ref{fig:resultspaths} for two different paths. Path $a$ corresponds to the center nodes located along the length of the beam, while path $b$  corresponds to the center nodes along the width direction.

In overall, we observe a very good agreement of the results. The differences observed in Fig.~\ref{fig:resultspaths} can be explained by the fact that we are comparing two different approaches for the resolution of unilateral contact problems. These differences can be attributed to the different formulations used in  \verb|Alya| and \verb|Code_Aster| for finite strains.

\begin{figure*}[h!]
\begin{center}
\includegraphics[width=12.0cm]{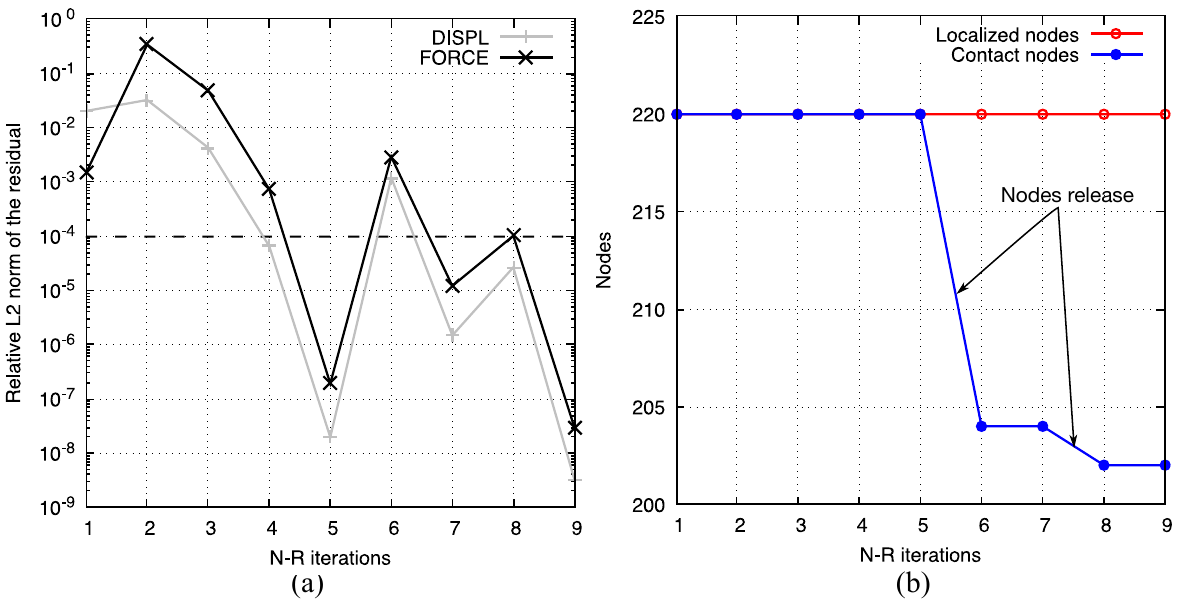} 
\caption{Convergence of the 3D indentation test for the time step No. 38. (a) convergence of the contact algorithm based on different convergence criteria and (b) subiterations for artificial nodes}
\label{fig:test-convergence}
\end{center}
\end{figure*}

The convergence of the contact algorithm is analyzed in Fig.~\ref{fig:test-convergence}. In this figure, we show the convergence of the Newton-Raphson (N-R) for the time step No. 38. This time step is selected as it allows to show how the nodes release is performed in an implicit time integration scheme (Algorithm~\ref{alg:reso}). Two convergence criteria from \cite{Belytschko2014} are evaluated: displacement increment error named as \verb|DISPL| and residual error, \verb|FORCE|, see Fig.~\ref{fig:test-convergence}a. The precision with which the error in displacement and force are calculated is set to \textbf{\SI{1e-4}{}} for both criteria. We can see a good convergence of the solution guaranteeing an equilibrium of the problem for each time step. When artificial contact nodes are detected, the contact algorithm requires subiterations to fulfill Eq.~\ref{eq:contactcondition}. Fig.~\ref{fig:test-convergence}b shows the additional iterations which are required within a time step to avoid the adhesion nodes at the contact region. The number of artificial (adhesion) nodes depends on the projection method and the geometry of the contact bodies and this is a key issue for further analysis.

\subsection{Drop-weight impact test}

\begin{figure}[h!]
\begin{center}
\includegraphics[width=8.0cm]{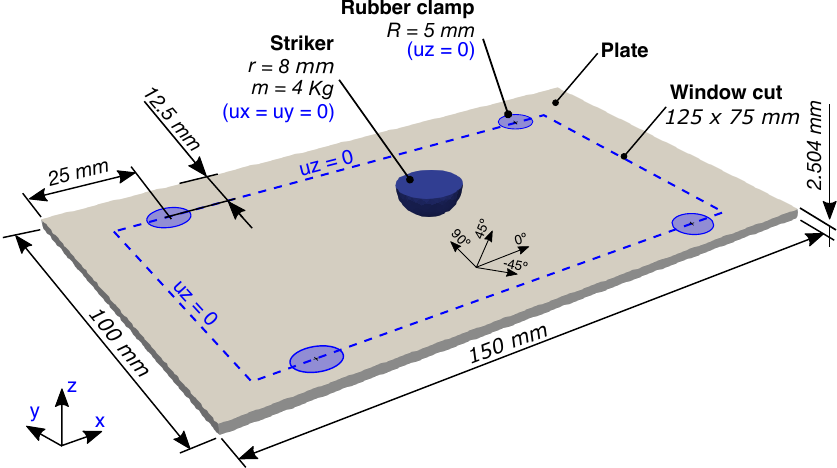} 
\caption{Drop-weight impact test model.}
\label{fig:lviproblem}
\end{center}
\end{figure}

In this example, we aim to solve a drop-weight impact test where an hemispherical rigid impactor is dropped in free fall on a plate, see Fig.~\ref{fig:lviproblem}. The impact setup and experimental results from \cite{Caputo2015} are used to compare the numerical predictions. In the present study the case of \SI{6}{\joule} (\SI{5.6}{\joule} according to the experiment)  is evaluated. In \cite{Caputo2015} the striker mass and its radius are not specified by the authors, so we assume $m=\,$\SI{4}{\kg} and $r=\,$\SI{8}{\mm}. The coupon is made of composite material with a stacking sequence of $[45/-45/90/0]_{s}$ and its dimensions are \qtyproduct{150 x 100 x 2.504}{\milli\metre}. The material is modeled using an orthotropic linear elastic model and its properties are summarized in Tab.~\ref{tab:matproplvi}.

\begin{table}[h!] 
\centering
\small
\begin{tabular}{llr}
\hline  
Property  & Description & Value \\
\hline  
$E_{11}$ (GPa)                  & Longitudinal Young modulus & $86.70$  \\
$E_{22}=E_{33}$  (GPa)   & Transverse Young modulus & $7.00$  \\
$\nu_{12}=\nu_{13}$ (-)   & Poisson ratio in 12 and 13 directions & $0.32$   \\
$\nu_{23}$  (-)                  & Out-of-plane Poisson ratio  &  $0.49$\\
$G_{12}=G_{13}$ (GPa)    & Shear modulus in 12 and 13 directions & $3.60$  \\
$G_{23}$  (GPa)                & Out-of-plane shear modulus  &$2.70$ \\
\hline
\end{tabular}
\caption{Elastic unidirectional lamina material properties \cite{Caputo2015}}
\label{tab:matproplvi}
\end{table}

According to the ASTM D7136/D7136M-20 standard \cite{Astmd7136}, the plate is supported on a metallic frame with a cut-out of \qtyproduct{125 x 75}{\milli\metre} and it is clamped at the four corners by rubber-tipped clamps. Equivalent boundary conditions are used for the model, see Fig.~\ref{fig:lviproblem}. With regards to the mesh, a refined centered region of \qtyproduct{75 x 75}{\milli\metre} is created with an element size of \SI{0.5}{\mm} in order to accurately capture the contact between impactor and plate. The global element size is \SI{1}{\mm} and each ply from the laminate stacking sequence has one element through the thickness. This results a base mesh of \num{324264} hexahedron elements ($\approx$  1.1 million of dof) for the plate. On the other hand, the impactor is meshed with 4-node linear tetrahedron elements with a biased mesh of \SI{0.184}{\mm} at the center of the half-sphere and \SI{1}{\mm} at the end of the edge. The total number of elements of the impactor base mesh is \num{8561} (\num{25683} dof). The total simulation time is set to \SI{8}{\milli\second} with a fixed time step size of \textbf{\SI{1e-5}{\second}}.

\begin{figure}[h]
\begin{center}
\includegraphics[width=7cm]{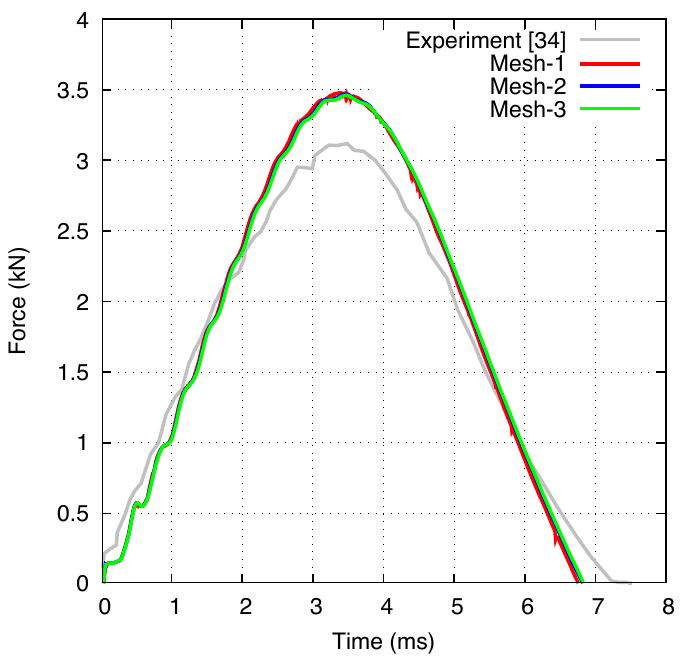} 
\caption{Mesh convergence for the 6J low velocity impact test. The reader is referred to the web version of this paper for the color representation of this figure.}
\label{fig:resultsimpactmesh}
\end{center}
\end{figure}

A mesh convergence analysis is conducted for this test, Tab.~\ref{tab:resmshes} summarizes three different meshes ranging from $324$K to $20$M of elements (plate instance). For the large case (Mesh-3) a total of \num{2400} CPUs are used to conduct the impact simulation. The force-time curve between the different meshes is shown in Fig.~\ref{fig:resultsimpactmesh}. As we can see in this figure, an excellent agreement is obtained between the different meshes demonstrating that the physical solution does not change when the mesh is refined and the number of processors is increased. On the other hand, comparing the numerical solution with the experimental result, we can see that the maximum impactor force is over-predicted respect to the experiment, a relative error of less than \SI{11}{\percent} is obtained respect to the experiment. With respect to the maximum indentations an error below \SI{1}{\percent} is obtained. The difference obtained for the maximum force can be explained due to the use of a fully elastic material model instead of specifically damage model for this kind of material. According to the authors in \cite{Caputo2015}, the material dissipates a small portion of energy due to delamination between plies, which is not taken into account in the present model. The use of proper advanced damage models for the prediction of onset and propagation of inter- and intra-laminar damage (delamination and fiber/matrix failure) it is required to have more accurate results from the material point of view. 

\begin{table*}[h]
\centering
\small
\begin{tabular}{lrrrrrrrr}
\hline  
Case  & No. elem. & No. CPUs & No. elem.  & No. CPUs & Max. force & Error & Max. displ. & Error \\
          & plate         &                   & impactor         &                  &(kN)          &  (\%)      & (mm)           & (\%)  \\
\hline
Mesh-1    &      \num{324264} &    \num{520}   &      \num{8561} &   \num{8}  &   3.480  & 10.24 & 3.800 & 0.07\\
Mesh-2   &    \num{5188224}  &    \num{742}   &   \num{68488} &  \num{16}  &   3.471  &  9.97  & 3.818 & 0.53 \\
Mesh-3   &  \num{20752896} &  \num{2304}  & \num{547904}  & \num{96} &   3.461   &  9.66  & 3.829 & 0.83 \\
\hline  
Experiment \cite{Caputo2015} &  &  & &  & 3.157 &  & 3.798 & \\
\hline
\end{tabular}
\caption{Comparison between numerical and experimental data for the 6J impact using different meshes.}
\label{tab:resmshes}
\end{table*}

\begin{figure*}[h!]
\begin{center}
\includegraphics[width=12cm]{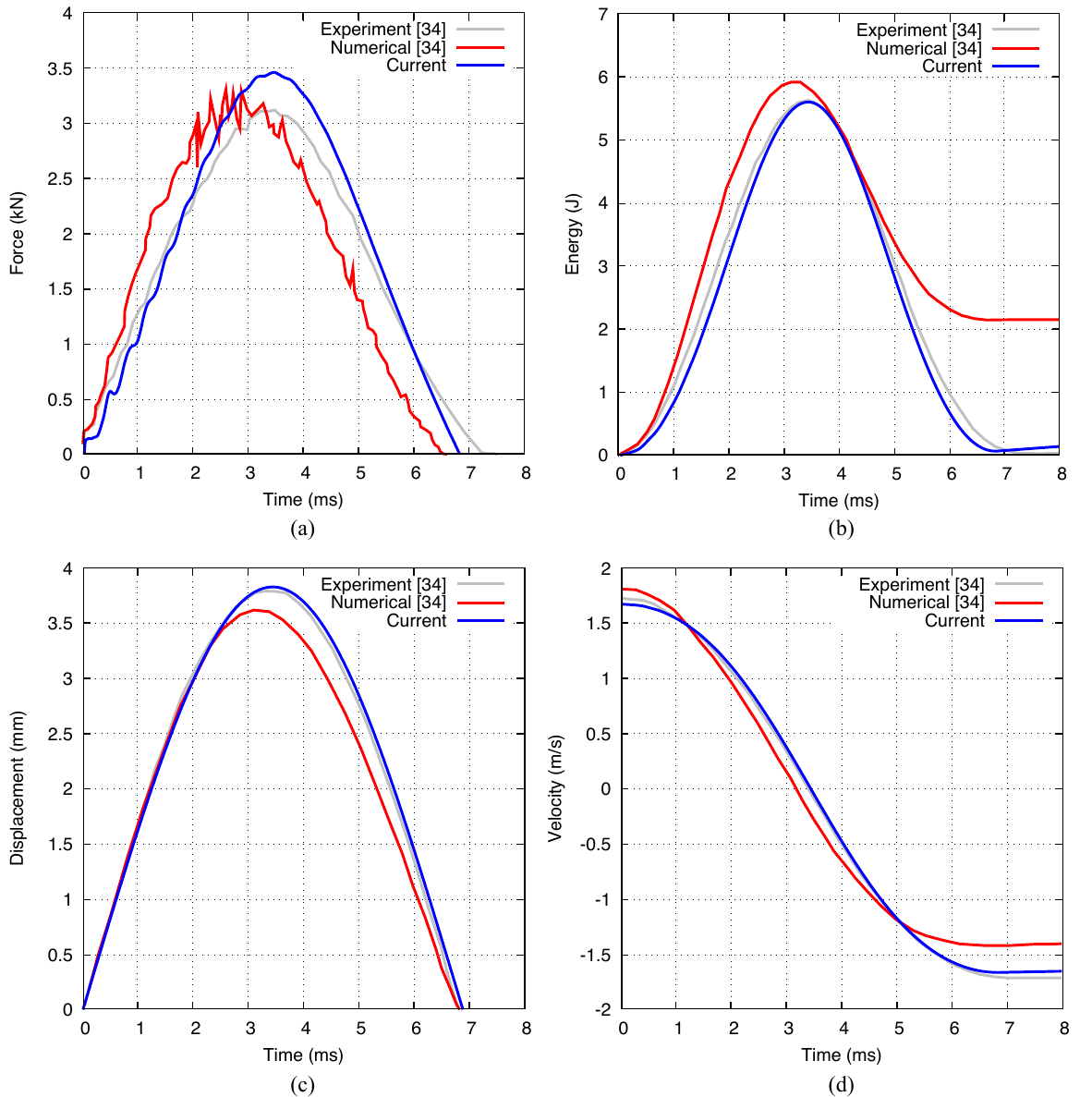} 
\caption{6J low velocity impact test (Mesh-3): (a) Impactor force vs. time. (b) Energy vs. time. (c) Impactor displacement vs. time. (d) Impactor velocity vs. time. The reader is referred to the web version of this paper for the color representation of this figure.}
\label{fig:resultsimpact}
\end{center}
\end{figure*}

Force, energy, displacement and velocity of the impactor versus time of impact are shown in Fig.~\ref{fig:resultsimpact}. The experimental result and the numerical solution obtained by the authors in \cite{Caputo2015} are also included in the figures for comparison. As we can see, a very good correlation is obtained for the displacement of the impactor (indentation), velocity and energy curves. These parameters demonstrate that the contact behavior and kinematics are well predicted by the proposed algorithm.

\section{Parallel performance}\label{sec:parallperformance}

In this section, we show the parallel performance of the proposed contact algorithm through the 3D indentation and the drop-weight impact tests previously discussed. In the first example (3D indentation test), we perform a trace analysis using a reduced number of processors in order to evaluate the behavior of the contact algorithm, while in the second example (drop-weight impact test) we evaluate the scalability of the code up to 2400 processors using a refined mesh of 80M elements. 

\paragraph*{\textbf{Trace analysis on the 3D indentation test}}
Trace executions are one of the main solutions for measurement and analysis of program performance on parallel computers. Traces are basically space-time diagrams that show how a parallel execution unfolds over time. They are analyzed {\it post-mortem}, as they are built based on the information gathered during the program execution. In a trace, time lines for different Message Passing Interface (MPI) processes are stacked top to bottom. A MPI process activity over time unfolds left to right. Each time line is composed of several colored segments, where each distinct color represent a different procedure, function or subroutine. Space-time visualization of MPI execution traces are very useful for spotting and understanding the temporal behaviour of the program, as they allow to graphically observe the work load balance among the different processors, the idle time for each processor, the time consumed for the execution of a given subroutine, the impact of inter-process communications, etc. In the present paper, we use the \verb|HPCToolkit|~\cite{hpctoolkit} suite in order to perform the trace analysis. 

\begin{figure}[h!]
\begin{center}
\includegraphics[width=8.5cm]{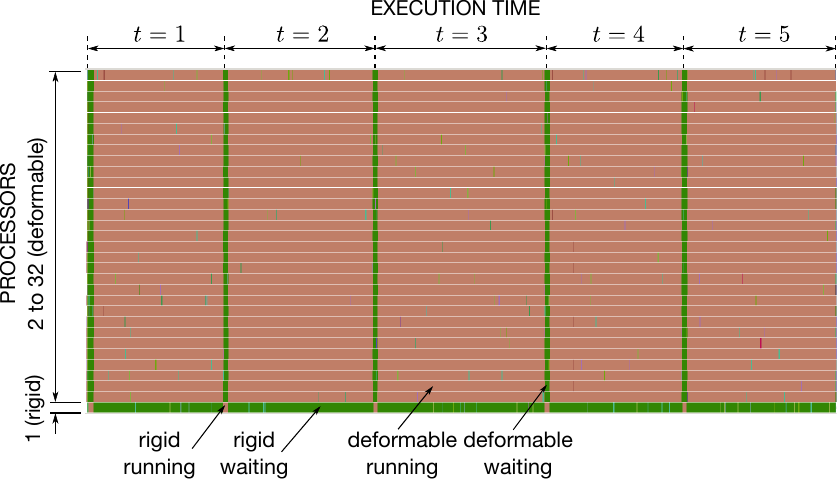} 
\caption{Global view of the execution trace for the 3D indentation test using 31 tasks for the beam and 1 task for the indenter \cite{Rivero2018PhD}. The reader is referred to the web version of this paper for the color representation of this figure.}
\label{fig:unilat_trace_all}
\end{center}
\end{figure}

A trace analysis is performed on the 3D indentation test, which has been previously described. The total number of processors used for the simulation is 32 (31 for the beam and 1 for the indenter). The execution trace obtained for five steps is depicted in Fig.~\ref{fig:unilat_trace_all}. The vertical axis of the trace corresponds to the processors used for each instance, while the horizontal axis is the space-time of the simulation. The orange color segments represent a running processor, while the green color means that the processor is idle or executing another task non-related to the solution of the finite element problem e.g., the contact detection. As we can see, the solution for the rigid body is faster than the deformable body. This is due to the complexity of solving the physics for each instance, e.g. the non-linearity of the deformable body. The block/segregated execution observed in the trace (first the rigid body, then the deformable body) is the Gauss-Seidel strategy adopted for the contact algorithm, which forces the processor dedicated to the rigid indenter to be idle when the remaining processors dedicated to the deformable beam are computing, and vice-versa. From the observation of this trace, we can also deduce that the workload is well balanced among the group of processors which belong to the deformable beam, as they start and finish its execution in a coordinated way. This means that all the computational resources are being used efficiently, as there is no waiting or idle time between processors when the deformable beam is computing. 

\begin{figure}[h!]
\begin{center}
\includegraphics[width=8.5cm]{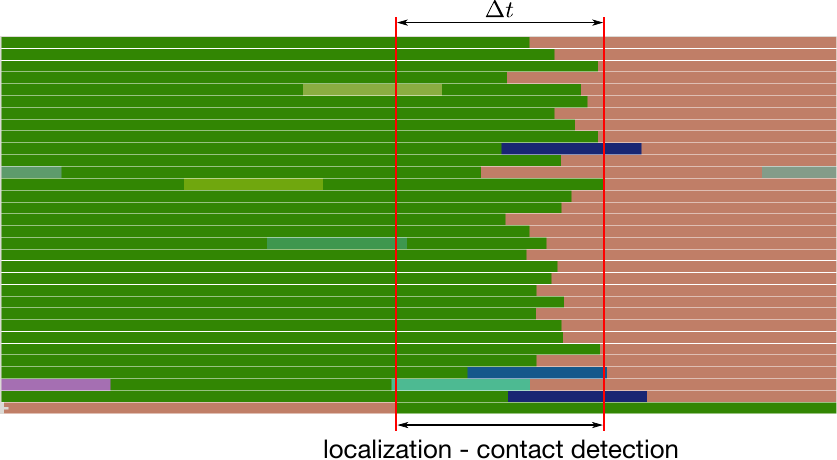} 
\caption{Local view (zoom in) on the time interval where localization occurs \cite{Rivero2018PhD}. The reader is referred to the web version of this paper for the color representation of this figure.}
\label{fig:unilat_trace_zoom}
\end{center}
\end{figure}

In order to analyze the localization and contact detection performed by \verb|PLE++| a zoom in the time interval is shown in Fig.~\ref{fig:unilat_trace_zoom}. As we can observe 
for this particular example, the localization task has a small impact in the overall execution time. The fact that the localization has a negligible impact in the total simulation time cannot be generalized without further analysis and it should be restricted only to the context of this particular example, as this impact strongly depends on the number of used processors. An increase in the number of processors involved in the localization implies more communications and operations, which could affect the execution time of the localization process. How the localization time affects the overall simulation time as the number of processors increase is a key issue for further analysis. 

\begin{figure}[h!]
\begin{center}
\includegraphics[width=8.0cm]{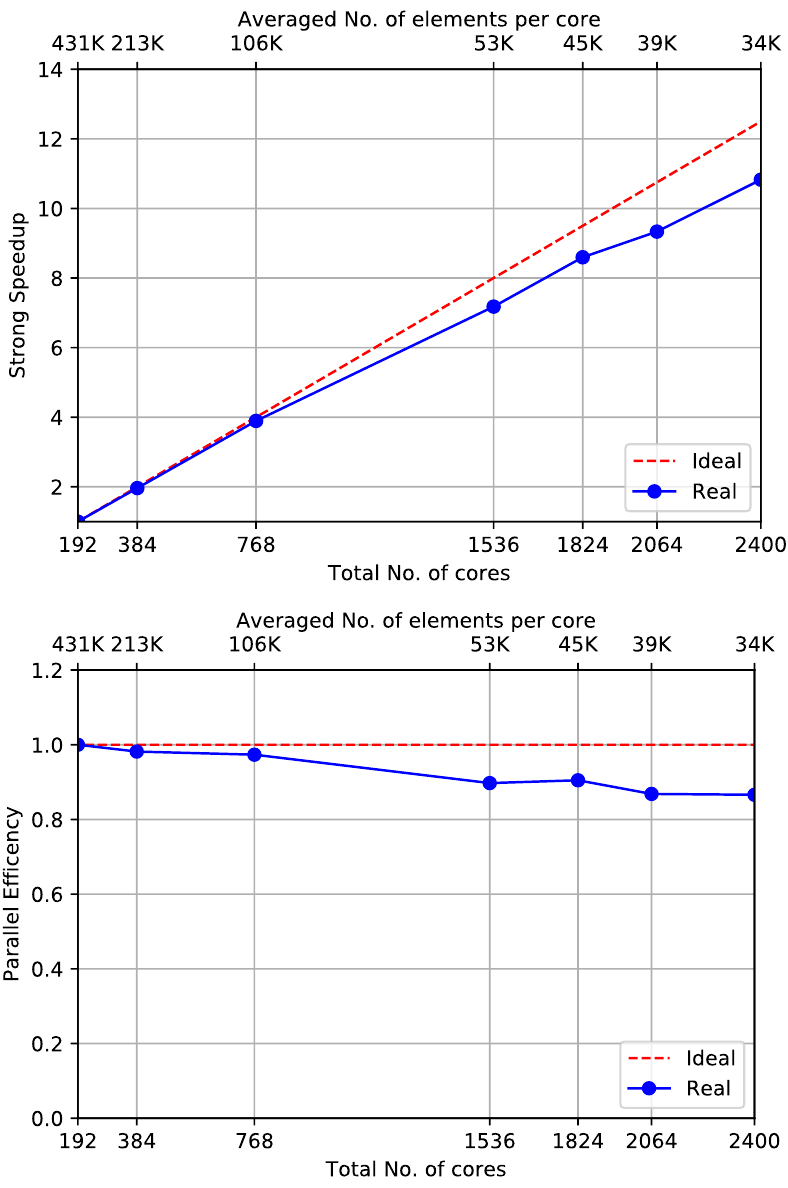} 
\caption{Strong scalability of the drop-weight impact test using a mesh of 80M of elements.}
\label{fig:scal80M}
\end{center}
\end{figure}

\paragraph*{\textbf{Scalability analysis on the drop-weight impact test}}
In the previous trace analysis, we could observe how the Gauss-Seidel method works in a "serial
manner" (while one of the bodies is computing the other is waiting). It is well-known that this
method is a-priori not the best option for efficient parallel resolutions but it is considered as a robust
method to prevent any numerical instabilities. However, it should be pointed out that there exist strategies
to fully exploit the resources in the case of the Gauss-Seidel method. In
\cite{doi:10.1080/10618562.2020.1783440}, an oversubscription is used for the different code
instances in the case of multiphysics problems. Such strategy is not envisaged in the present
work as the rigid body simulation only requires few cores to run, meaning that the unused resources
are limited.
To evaluate the speedup and the parallel efficiency of the proposed approach strong scalability has been conducted on the drop-weight impact test previously described. In this case, a larger mesh than the one used for the physics validation has been evaluated with 80M of elements ($\approx$ 250 million of dof). Strong scalability consists of fixing the mesh and solving the problem with a different number of processors. The strong speedup is calculated as $\frac{t_{0}}{t_{N}}$ while the parallel efficiency is calculated as $\frac{t_{0}N_{0}}{t_{N}N}$, where $N$ is the number of cores and $t_{0}$ is the reference simulation time for $N_{0}$ cores. The number of processors used ranges from 192 to \num{2400}. As we are working with a multibody and multicode approach, we fix the number of processors for the rigid body (impactor) to 3 processors (sufficiently for its base mesh of \num{8561} elements) and we change the number of processors of the deformable body where we have the strong computational effort. Only the first 15 time steps are analyzed, which are sufficient as contact is occurring between the impactor tip and the plate during this time. Strong speedup and parallel efficiency is depicted in Fig.~\ref{fig:scal80M}. The ideal scalability and efficiency are represented with a dashed line.

We can see that the scalability of the problem is really good up to \num{2400} processors, maintaining a parallel efficiency above 80\%. Moreover, the optimum number of elements per core according to the aforementioned efficiency is around 34K elements. Therefore, this analysis demonstrates the scalability of the proposed algorithm and also provides a good indication of the amount of resources to request for the size of a particular problem.
 
\section{Concluding remarks}

The development of a full parallel contact algorithm to solve rigid-to-deformable bodies is presented. The approach is based on partial Dirichlet-Neumann boundary conditions and it works as a Gauss-Seidel scheme with weak coupling. The main aspect of the proposed algorithm is that the bodies are treated separately, in a segregated way. As this approach solves each body separately (multicode), there is no need to increase the degrees of freedom of the problem or to redefine the mesh graph at different time steps, since no contact elements are used in the algorithm. Depending on the complexity of the problem, different time integration schemes can be used in the contact resolution to solve the physics for each body.

The verification and validation of the algorithm are conducted in several benchmark cases. In all the cases, a good agreement is obtained between the proposed solution and other solutions obtained analytically, numerically (from other finite element codes) or experimentally, demonstrating the capabilities of the proposed algorithm for 2D and 3D problems. The last example corresponds to a well-known test named as drop-weight impact test, which is used by the industry for the damage resistance of materials. In this case, the proposed algorithm is validated through experiments showing its reliability in real problems. Moreover, a mesh convergence analysis is conducted for the drop-weight impact test showing that the physics of the problem remains the same by using different meshes of the model and increasing the number of processors. The refined mesh is discretized with 20M of elements and solved with \num{2400} processors showing its applicability for large-scale problems.

In terms of parallel performance, a trace analysis and a strong scalability analysis are conducted for the 3D indentation test and the drop-weight impact test respectively. For the case of the 3D indentation test, we show that the workload related to the overall execution including the localization/detection phase is well balanced among the group of processors assigned to each body. This means that all the computational resources are being used efficiently, as there is no waiting or idle time between processors, in particular when the deformable body is computing. On the other hand, a strong scalability analysis is conducted on the drop-weight impact test with a mesh of 80M of elements. This case shows a very good parallelization efficiency above 80\% up to 2400 processors and demonstrates its capability for larger problems. However, the application of this contact algorithm in other case scenarios such as contact between bodies of similar size would be a key issue for further computational performance analysis.  

A further conclusion from this work is that many engineering contact problems involve the contact between rigid and deformable bodies such as indentation tests, impact tests, metal forming, tire-road contact or even biomechanical applications such as a stent in arteries. In all theses cases, the proposed method can be very useful thanks to its simplicity, robustness of the fulfillment of the geometrical constraints and efficiency for large-scale problems requiring a high computational cost.

\section{Acknowledgements}

This is work is an extension of the PhD thesis performed by M. Rivero, who specially acknowledges the financial support for the predoctoral grant BES-2012-052278 from the Spanish Government and the Severo Ochoa Programme SEV-20110067. This work has also received funding from the Clean Sky 2 Joint Undertaking (JU) under grant agreements No. 807083 and No. 945521. The JU receives support from the European Union’s Horizon 2020 research and innovation programme and the Clean Sky 2 JU members other than the Union. Moreover, the author thankfully acknowledges the computer resources at MareNostrum and the technical support provided by Barcelona Supercomputing Center (RES-FI-2018-3-0022).

\bibliographystyle{elsarticle-num}
\bibliography{Guillamet2022a.bib}

\end{document}